\documentclass[aps,prb,reprint,superscriptaddress]{revtex4-1}
\usepackage{graphicx}
\usepackage{color}
\usepackage{amsmath}

\begin{document}


\title{Noncoplanar ferrimagnetism and local crystalline-electric-field anisotropy in the quasicrystal approximant Au$_{70}$Si$_{17}$Tb$_{13}$}



\author{Takanobu Hiroto}
\affiliation{National Institute for Materials Science, Tsukuba, Ibaraki 305-0047, Japan}

\author{Taku J Sato}
\email[]{taku@tohoku.ac.jp}
\affiliation{Institute of Multidisciplinary Research for Advanced Materials, Tohoku University,
  2-1-1 Katahira, Aoba, Sendai 980-8577, Japan}

\author{Huibo Cao}
\affiliation{Neutron Scattering Division, Oak Ridge National Laboratory, Oak Ridge, Tennessee 37831}

\author{Takafumi Hawai}
\author{Tetsuya Yokoo}
\author{Shinichi Itoh}
\affiliation{Institute of Materials Structure Science, High Energy Accelerator Research Organization (KEK), Tsukuba, Ibaraki 305-0801, Japan}

\author{Ryuji Tamura}
\affiliation{Department of Materials Science and Technology, Tokyo University of Science,  Katsushika, Tokyo 125-8585, Japan}


\date{\today}

\begin{abstract}
  Neutron scattering experiments have been performed to elucidate magnetic properties of the quasicrystal approximant Au$_{70}$Si$_{17}$Tb$_{13}$, consisting of icosahedral spin clusters in a body-centered-cubic lattice.
  Bulk magnetic measurements performed on the single crystalline sample unambiguously confirm long-range ordering at $T_{\rm C} = 11.6 \pm 1$~K.
  In contrast to the simple ferromagnetic response in the bulk measurements, single crystal neutron diffraction confirms a formation of intriguing non-collinear and non-coplanar magnetic order.
  The magnetic moment direction was found to be nearly tangential to the icosahedral cluster surface in the local mirror plane, which is quite similar to that recently found in the antiferromagnetic quasicrystal approximant Au$_{72}$Al$_{14}$Tb$_{14}$.
  Inelastic neutron scattering on the powdered sample exhibits a very broad peak centered at $\hbar \omega \simeq 4$~meV.
  The observed inelastic spectrum was explained by the crystalline-electric-field model taking account of the chemical disorder at the fractional Au/Si sites.
  The resulting averaged anisotropy axis for the crystalline-electric-field ground state is consistent with the ordered moment direction determined in the magnetic structure analysis, confirming that the non-coplanar magnetic order is stabilized by the local uniaxial anisotropy.
\end{abstract}


\maketitle

\section{Introduction}
Icosahedral quasicrystals are solids that exhibit sharp Bragg reflections with the icosahedral rotational symmetry, first observed in an electron diffractogram~\cite{ShechtmanD84}.
It has been well known that the icosahedral rotational symmetry is incompatible with the lattice periodicity; indeed, atomic structure of the icosahedral quasicrystals is described as a quasiperiodic packing of icosahedral atomic clusters~\cite{TakakuraH07}.
When the bulk icosahedral symmetry is relaxed, the atomic clusters may form a paeriodic lattice, by deforming the cluster slightly and/or introducing glue atoms between the clusters.
Indeed, such approximant crystals have been known to form in the vicinity of icosahedral quasicrystal phases in alloy phase diagrams; several detailed studies on atomic structures confirmed that the approximant crystals are made of quite similar atomic clusters to those in the icosahedral quasicrystals~\cite{GoldmanAI93,TakakuraH07}.

Among various quasicrystals and approximants ever found, the icosahedral Cd-Yb quasicrystal and related approximants opened a new era for the research of the magnetism of quasicrystals.
These quasicrystals and approximants consist of the so-called Tsai-type icosahedral clusters, which are made of four matryoshka-like successive shells~\cite{Gomez2003, PiaoSY06}; the first shell is a dodecahedron composed of 20 Cd atoms, the second shell is an icosahedron of 12 rare-earth ($R$) atoms, the third shell is an icosidodecahedron of 30 Cd atoms, and the outermost shell is a defect rhombic triacontahedron of 84 Cd atoms.
To date, the Tsai-type icosahedral clusters have been identified in a number of approximants, e.g., Cd-$R$~\cite{BruzzoneG71,TsaiAP00, GoldmanAI13}, (Cd, Zn)-Mg-$R$~\cite{GuoJQ00,GuoJQ01,GuoJQ02,TsaiAP94}, Ag-In-$R$~\cite{MoritaY08}, and Au-$SM$-$R$ ($SM$ = Si, Ge, and Sn)~\cite{MoritaY08,KenzariS08,GebresenbutGH13}.
The unique feature of the Tsai-type clusters is that the second icosahedral shell is exclusively occupied by the $R$ elements [see Fig.~1(a)], realizing a defectless network of magnetic moments.
It should be noted, however, that the cluster center is occupied by a Cd$_4$ tetrahedron that is in most cases orientationally disordered at room temperature~\cite{Tamura2002,TamuraR05,NishimotoK13}.
In addition, for the ternary and quaternary systems, there are fractional sites occupied by the elements other than $R$, and hence the randomness due to the chemical disorder exists.
Nonetheless, effect of such disorder is presumably much weaker compared to the disorder in the magnetic-moment network itself, and hence the Tsai-type quasicrystals and approximants are believed to be suitable to study magnetism of quasiperiodic and periodic arrays of the icosahedral spin clusters.

While long-range magnetic orders have not yet been observed in the Tsai-type quasicrystals, various magnetic orders, i.e., antiferro-, ferri-, and ferro-magnetic orders, have been observed in Cd$_6$$R$ ($R$ = Nd, Sm, Gd, Tb, Dy, Ho, Er, and Tm)~\cite{Tamura2010,MoriA12,TamuraR12} and Au-$SM$-$R$ ($SM$ = Si, Ge, and Sn; $R$ = Gd, Tb, Dy, and Ho)~\cite{Hiroto2013,HirotoT14} approximant crystals.
Most of them have been identified by bulk magnetic measurements, whereas a few were investigated by synchrotron X-ray magnetic diffraction experiments~\cite{Kim2012,KreyssigA13}.
Microscopic magnetic-structure analysis in their long-range-ordered state has been rather limited; only ferromagnetic (ferrimagnetic) Au-Si-$R$ ($R =$ Tb and Ho) and antiferromagnetic Au-Al-Tb approximants were investigated to date.
For Au-Si-Tb, the ferrimagnetic-like collinear magnetic structure was proposed based on a powder neutron diffraction experiment~\cite{Gebresenbut2014}.
According to their model, the Tb$^{3+}$ moment sizes substantially vary from 1.7 to 8.2~$\mu_{\rm B}$, although all twelve Tb atoms on the icosahedron belong to a single crystallographic site (24g).
Very recently, a preliminary result has been reported on the magnetic structure analysis of the Au-Si-Ho and Au-Si-Tb approximants.
A somewhat non-collinear magnetic structure was proposed with moments mostly orthogonal to each other~\cite{GebresenbutGH19}, nonetheless details are not yet published to date.
Phenomenological theoretical investigation was also performed on a possible non-collinear order in the Au-Si-Tb approximant, in which effect of the single ion anisotoropy was emphasized~\cite{SugimotoT16}.

For the antiferromagnetic quasicrystal approximant Au$_{72}$Al$_{14}$Tb$_{14}$, we have recently determined the magnetic structure using neutron diffraction with the aid of the magnetic representation analysis~\cite{SatoTJ19}.
The magnetic structure was found to be made of non-coplanar whirling spins on the icosahedral spin cluster, arranged in an antiferroic manner breaking the body-centered-cubic (bcc) translational invariance.
It was suggested that the ordered spin directions were primarily fixed by the local anisotropy reflecting the symmetry of the icosahedral cluster.
It is rather puzzling to see very different magnetic ordering in the two seemingly similar alloy systems, Au-Si-Tb and Au-Al-Tb, and hence, it may be worthwhile to revisit the magnetic structure of the Au-Si-Tb quasicrystal approximant.

Related to the local anisotropy, another key information to understand the magnetic structure is the details of the crystalline-electric-field (CEF) ground state for 4f elections of the $R^{3+}$ ions.
The CEF splitting in the Tsai-type cluster compounds was investigated by bulk measurements in the Zn-Ag-Sc-Tm quasicrystal and Zn-Sc-Tm approximant~\cite{JazbecS16}, as well as by neutron inelastic scattering in the Cd$_{6}$Tb approximant~\cite{DasP2017}.
Both the studies infer the dominant second order uniaxial $B^0_2$ term in the CEF Hamiltonian, whereas the former suggests existence of additional weak pseudo five-fold ($5f$) term $B^5_6$.
This result consequently infers that the uniaxial anisotropy axis of the rare-earth ions is along the pseudo $5f$-axis of the icosahedral cluster, and is apparently inconsistent with the results of the magnetic structure determinations.

As briefly summarized in the above, the microscopic arrangement of spins in the ordered phase, as well as the details of CEF anisotropy, is still controversial.
This may be due to the fact that there is no complete study of magnetic structure and excitations in one quasicrystal approximant system.
We, therefore, undertook thorough elastic and inelastic neutron scattering study on the quasicrystal approximant Au$_{70}$Si$_{17}$Tb$_{13}$.
The neutron diffraction study using a high-quality single crystalline sample clearly shows that the magnetic order is non-collinear and non-coplanar, and is closely related to the whirling spin order observed in the antiferromagnetic counterpart Au$_{72}$Al$_{14}$Tb$_{14}$.
The neutron inelastic scattering shows existence of a broad excitation peak around $\hbar \omega \simeq 4$~meV, which is attributable to the excitation between the CEF splitting levels.
The point-charge calculation taking account of the chemical disorder successfully reproduces the broadness of the excitation peak.
The average moment direction obtained in the point-charge calculation is nearly perpendicular to the local pseudo $5f$-axis, and is in good accordance with the observed moment direction.
By combining both the magnetic structure and excitation results, we conclude that the dominant term in the CEF Hamiltonian is $B^0_2$, as inferred in the earlier studies, nonetheless, the quantization (anisotropy) axis is nearly perpendicular to the pseudo $5f$-axis in the present Au$_{70}$Si$_{17}$Tb$_{13}$ approximant.

\section{Experimental}

\begin{figure}
  \includegraphics[scale=0.35, angle=-90, trim= 0 60 0 0 ]{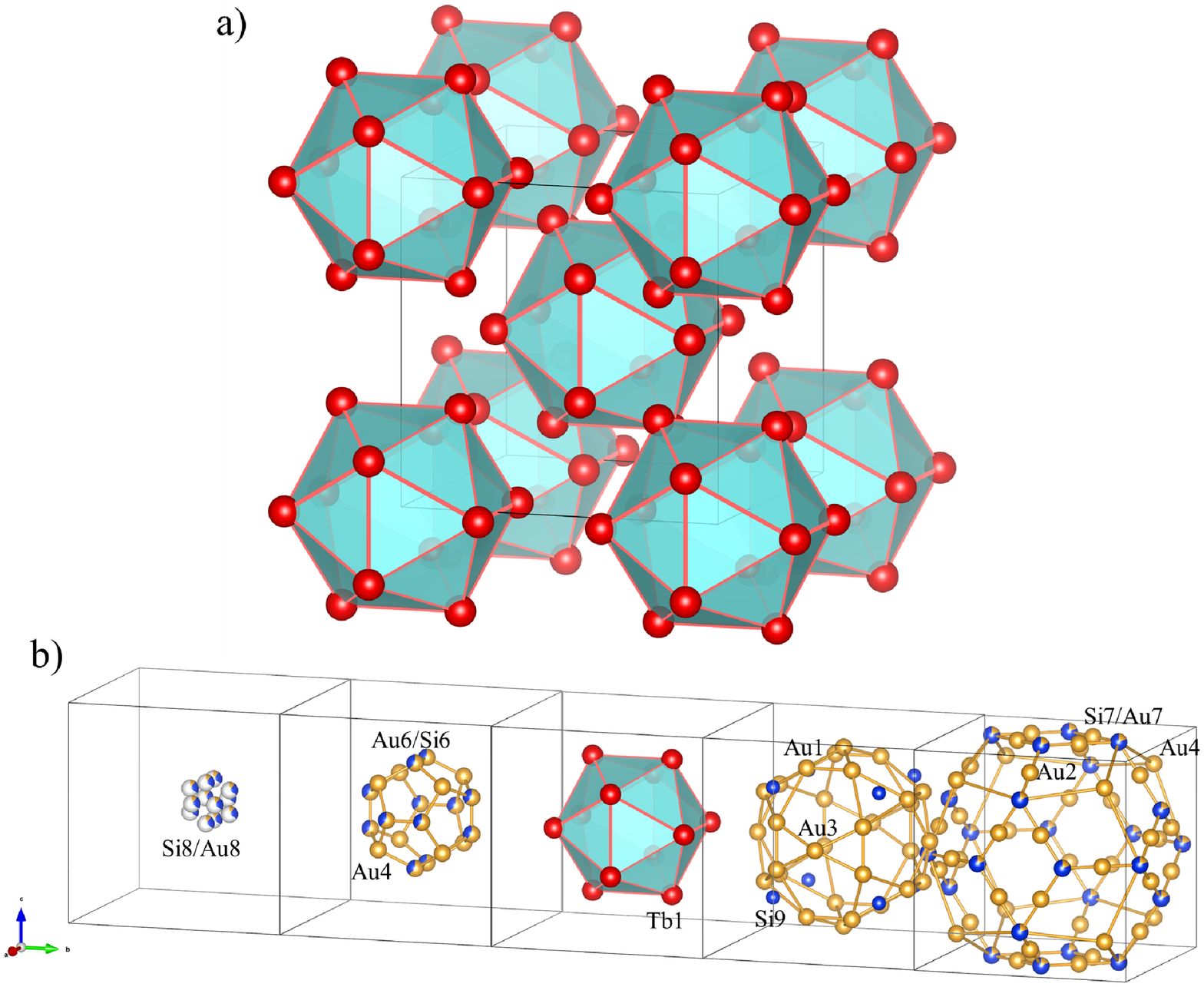}
  \caption{\label{figure1} (a) Body-centered-cubic array of Tsai-type icosahedral clusters in the Au-Al-Tb approximant.
    (b) Multiple shell structure of the Tsai-type cluster.
    Magnetic Tb$^{3+}$ ions occupy the second shell, selectively.
  }
\end{figure}

High purity elements better than 99.9~wt.~\% of Au, Si, and Tb with a nominal composition of Au$_{74.9}$Si$_{17.1}$Tb$_{8}$ were placed in an alumina crucible sealed inside a quartz ampule under an argon atmosphere.
The ampule was placed inside an electric furnace and melted at 1373~K for 5~h, and then cooled to 923~K.
Single grains were grown by slow cooling of the melt to 823~K at a rate of 1~K/h.
At 823~K, the ampule was taken out from the furnace and the melt was removed by using a centrifuge.
The obtained single grains have well-defined \{100\} and \{110\} facets with sizes up to 2.5~mm.

The phase constitution was checked by the powder X-ray diffraction (XRD) using the CuK$\alpha$ radiation (Rigaku, MiniFlex).
The quality of the single grains was examined by transmission high-energy X-ray Laue method (YXLON MG452 x-ray generator operating at 450~kV/5~mA, equipped with a CCD camera).
The alloy composition was checked by the energy dispersive x-ray spectroscopy (EDX) measurement in the scanning electron microscope (JEOL, JEM-2010F), and resulting alloy composition was Au$_{69.5(5)}$Si$_{17.2(4)}$Tb$_{13.3(3)}$.
(We use an approximate chemical formula Au$_{70}$Si$_{17}$Tb$_{13}$ in the following for simplicity.)
The DC magnetization was measured using a superconducting quantum interference device (SQUID) magnetometer (Quantum Design, MPMS-XL) down to 2~K with the external magnetic field up to 50000~Oe.
A single crystal with dimensions of about $1.5 \times 1.5 \times 0.5$~mm$^3$ (16~mg) was used for the magnetization measurement.

The polycrystalline sample was synthesized from high purity ($> 99.9$~wt\%) elements Au, Si and Tb by arc-melting.
As-cast alloy was subsequently annealed at 973~K for 50~h under an Ar atmosphere to obtain a single-phase sample.
The phase purity and crystal structure were checked by the powder XRD using the CuK$\alpha$ radiation.
The magnetic susceptibility was checked in a similar manner as described above.

The single crystal neutron diffraction experiment was performed using the four-circle diffractometer HB-3A at the high flux isotope reactor (HFIR), Oak Ridge national laboratory (ORNL), USA.
Diffraction datasets were collected at 15~K (in the paramagnetic phase) and 5~K (in the ferromagnetic phase) with a constant wavelength of 1.542~\AA$^{-1}$ selected by a bent perfect Si 220 monochromator~\cite{ChakoumakosBC11}.
The temperature uncertainty during the experiments was less than 1~K.
Integrated intensity of nuclear and magnetic Bragg reflections was measured for $0 < h < 9$, $0 < k < 9$, $0 < l < 8$ in the reciprocal-lattice space ($2\theta < 61.5^{\circ}$). 
Total 90 (for nuclear) and 100 (for magnetic) reflections with $I > 2\sigma$ were used for the crystal-structure and magnetic-structure refinements, respectively.
For the magnetic structure refinement, the differences in the Bragg peak intensities between $T = 5$ and 15~K were used.
Basis vectors of the irreducible representations of the magnetic representation were obtained using the {\sf MSAS} program~\cite{MSAS2019}, and the least-square fitting was performed using the linear combination of the basis vectors as a trial spin structure.

The powder neutron inelastic scattering experiment has been performed using the high-resolution chopper (HRC) spectrometer, installed at the materials and life science facility (MLF), Japan proton accelerator research complex (J-PARC)~\cite{ItohS11}.
The inelastic scattering spectra were obtained mainly using the incident energies $E_{\rm i} = 13.29$~meV, whereas other higher incident energies were also used to see overall features of the excitation spectrum.
Typical energy resolution for $E_{\rm i} = 13.29$~meV was $\Delta E/E \simeq 3$~\% (full-width at half-maximum; FWHM) at the elastic position in the present setup.
The powdered sample was inserted in the double annular sample container to reduce the self attenuation effect.
The container was then sealed in a standard aluminum can with the He heat-exchange gas, and set to the cold head of the closed cycle $^4$He refrigerator with the lowest working temperature around 3~K.

\section{Results and Discussion}

\subsection{Structural refinement}

\begin{figure}
  \includegraphics[scale=0.3, angle=-90]{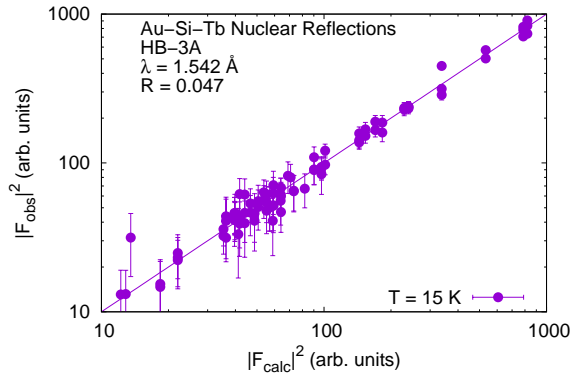}
  \caption{\label{figure2} $|F_{\rm obs}|^2$ versus $|F_{\rm calc}|^2$ for nuclear Bragg reflections.
    $|F_{\rm obs}|^2$ were collected at $T = 15$~K.
    The parameters used for the $|F_{\rm calc}|^2$ calculation are summarized in Table~\ref{table1}.
    The final conventional $R$-factor is 0.047, and weighted $\chi^2$ is 1.75.
  }
\end{figure}

First, we performed structural refinement of the Au$_{70}$Si$_{17}$Tb$_{13}$ compound in the paramagnetic region, i.e., at 15~K, using the reported model~\cite{GebresenbutGH16} as an initial structure.
The atomic positions and the occupancies of Au/Si fractional sites were refined simultaneously.
The resulting $|F_{\rm obs}|^2$ versus $|F_{\rm calc}|^2$ plot is given in Fig.~\ref{figure2}.
The final conventional reliability factor $R$ and weighted $\chi^2$ parameters are 0.047 and 1.75, respectively.
Table~\ref{table1} shows the refined structural parameters for the Au$_{70}$Si$_{17}$Tb$_{13}$ approximant obtained in the present study.
The refined parameters are mostly consistent with the one reported in the earlier study~\cite{GebresenbutGH16}.
It may be noted that the center of cluster was assumed to be empty in the above refinement; we have tried to incorporate finite Tb occupancy at the cluster center, however, it only degrades the fitting quality.
In the earlier work, it is reported that the occupancy of cluster-center Tb becomes finite as the Tb concentration increases from the minimum value of 13.63~\%.
Since the Tb concentration (14~\%) determined in our neutron diffraction has rather large uncertainty, we cannot compare the estimated composition with sufficiently high precision.
Nonetheless, since we grew the single crystal from the low Tb-concentration liquid (see experimental), we believe that our sample is at the lower-Tb-concentration edge, and hence should correspond to the 13.63~\% Tb concentration in the earlier work.
Indeed, the Tb concentration of the present single crystal determined by the EDX measurement is 13.3(3)~\%, which is rather close to the minimum Tb concentration of the earlier work.
Therefore, having this Tb composition in mind, the absence of the cluster-center Tb deduced in the present structure analysis is consistent with the earlier work.

\begin{table*}
  \caption{\label{table1} Refined structural parameters at $T = 15$~K for the Au$_{70}$Si$_{17}$Tb$_{13}$ approximant.
    The space group $Im\bar{3}$ were used in the refinement.
    The lattice constant is $a = 14.726$~\AA.
    Number of reflections used in the refinement was 90 ($I > 2\sigma$), whereas number of refined parameters was 20.
    Calculated chemical formula is Au$_{68}$Si$_{18}$Tb$_{14}$, and calculated density is 14.73~g/cm$^3$.
    The absorption coefficient for $\lambda = 1.542$~\AA\ is $\mu = 3.3$~cm$^{-1}$, however, since the sample was small and almost spherical, no absorption correction was made.
    Secondary extinction effect was corrected assuming exponential deviation from the calculated squared structure factor.
    The conventional $R$-factor is $R = 0.047$, whereas weighted $\chi^2 = 1.75$.
  }
  \begin{ruledtabular}
    \begin{tabular}{l c c c c c c}
      Site & Wyckoff position & $x$ & $y$ & $z$ & $B_{\rm iso}$ (\AA$^2$) & Occupancy \\ \hline 
      Au1 & 48h & 0.107(4) & 0.338(4) & 0.204(4) & 0.58(46) & 1        \\
      Au2 & 24g & 0        & 0.401(6) & 0.355(6) & 0.93     & 1       \\
      Au3 & 12d & 0.406(9) & 0        & 0        & 1.71     & 1       \\
      Au4 & 16f & 0.151(4) & 0.151    & 0.151    & 0.97     & 1       \\
      Au6 & 24g & 0        & 0.239(7) & 0.083(7) & 0.89     & 0.63    \\
      Si6 & 24g & 0        & 0.239    & 0.083    & 0.89     & 0.37(30)\\
      Au7 & 12e & 0.200(14)& 0        & 0.5      & 0.81     & 0.19    \\
      Si7 & 12e & 0.200    & 0        & 0.5      & 0.81     & 0.81(38)\\
      Au8 & 24g & 0        & 0.093(28)& 0.054(30)& 1.85     & 0.11    \\
      Si8 & 24g & 0        & 0.093    & 0.054    & 1.85     & 0.23(47)\\
      Si9 & 8c  & 0.25     & 0.25     & 0.25     & 0.91     & 1       \\
      Tb1 & 24g & 0        & 0.187(6) & 0.306(5) & 0.42     & 1       \\
    \end{tabular}
  \end{ruledtabular}
\end{table*}

\subsection{Magnetic susceptibility, magnetization and neutron diffraction intensity}

\begin{figure}
  \includegraphics[scale=0.45, angle=180, trim=0 0 0 0]{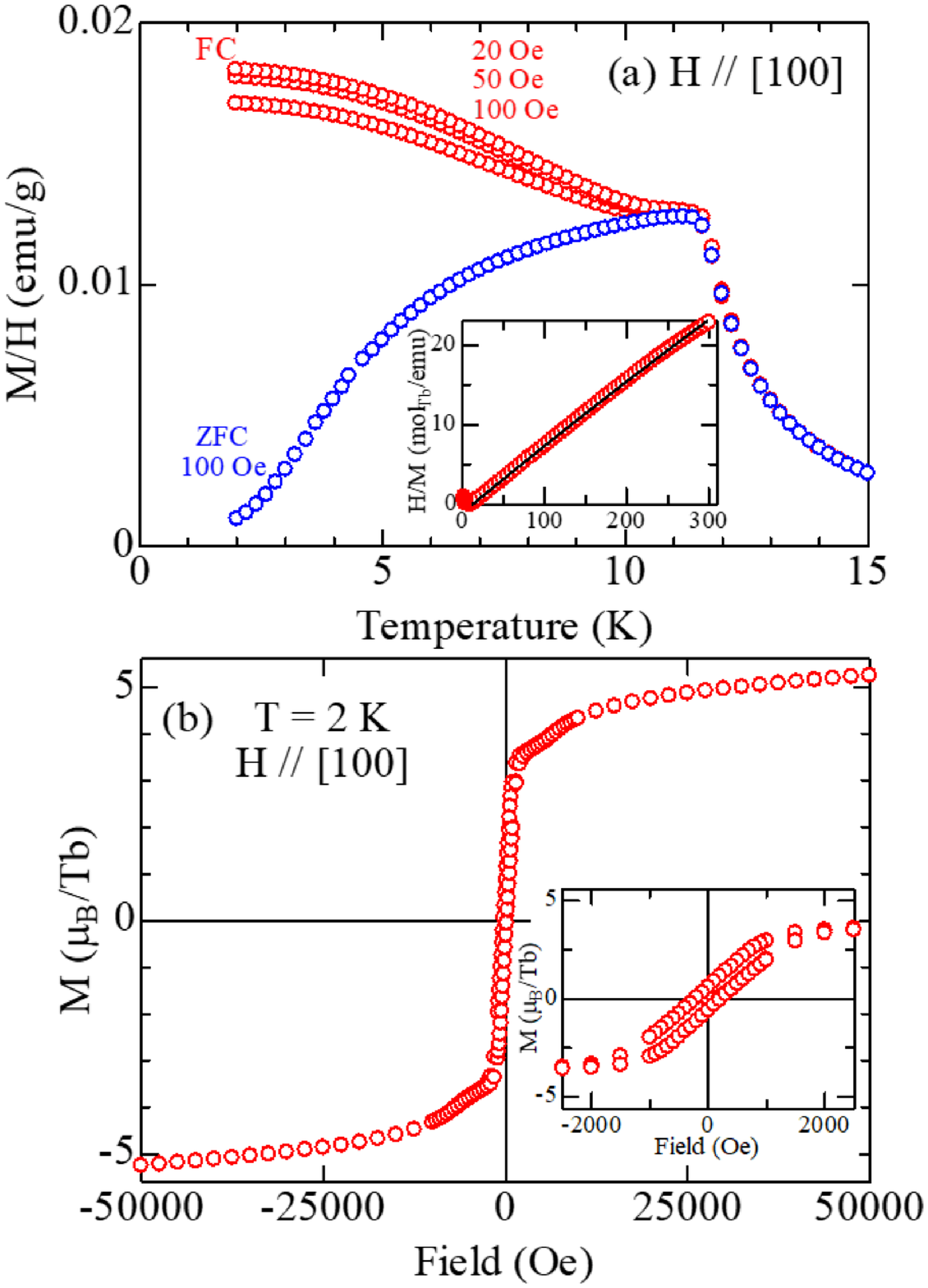}
  \caption{\label{figure3} (a) Temperature dependence of the magnetic susceptibility measured under the external magnetic field along the $[100]$ axis.
    For the field-cooled (FC) runs, data taken under $H = 20$, 50 and 100~Oe are shown, whereas only $H = 100$~Oe data are shown for the zero-field-cooled (ZFC) run.
    Inset: Inverse susceptibility for the wide temperature range up to 300~K.
    The solid line stands for the result of the Curie-Weiss fitting.
    (b) External magnetic field dependence of the magnetization ($M$-$H$ Curve) observed at $T = 2$~K.
    Magnetic field was applied along the $[100]$ direction.
    Inset: a magnified figure in the low-field region.}
\end{figure}

\begin{figure}
  \includegraphics[scale=0.3, angle=90, trim=0 50 0 0]{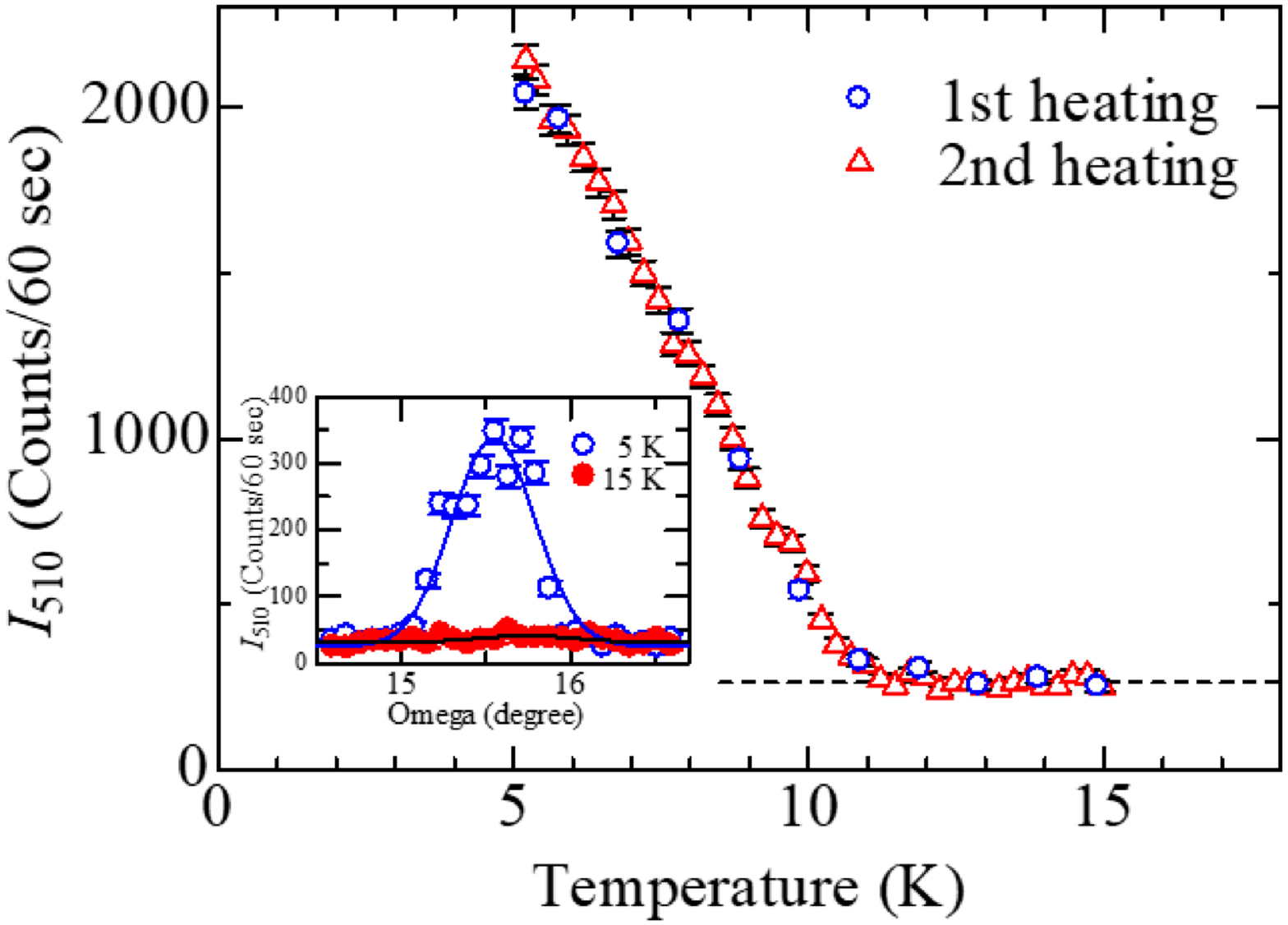}
  \caption{\label{figure4} Temperature dependence of the integrated intensity of the 510 Bragg reflection.
    Results of the two independent heating runs are shown.
    Inset: $\omega$-scans at the two temperatures $T = 5$ and 15~K around the 510 reflection position.}
\end{figure}

\begin{figure}
  \includegraphics[scale=0.4, angle=90, trim=100 150 100 100]{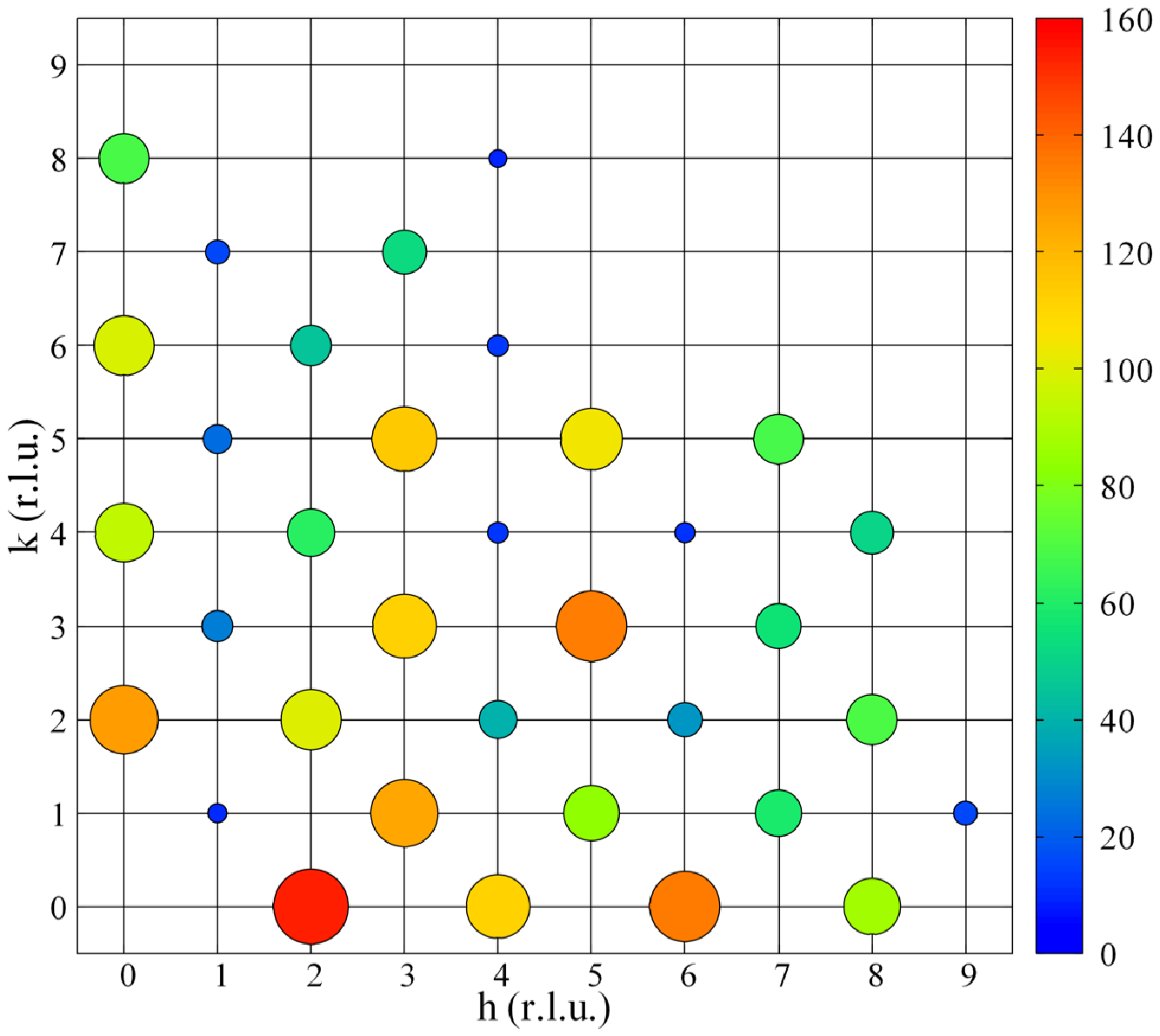}
  \caption{\label{figure5} Distribution of the observed magnetic structure factors $|F_{\rm obs}|$ (represented by radius of circles) of the magnetic Bragg reflections in the $hk0$ plane.
    The magnetic intensity was deduced by taking difference of the integrated reflection intensities between $T = 5$ and 15~K.}
\end{figure}

The field-cooled (FC) and zero-field-cooled (ZFC) magnetization was measured under various external magnetic field up to 100~Oe along $\vec{H} \parallel [100]$.
Fig.~\ref{figure3}(a) shows the resulting temperature dependence of the magnetic susceptibility ($M/H$), together with the inverse susceptibility in its inset. 
The inverse magnetic susceptibility shows apparent linear behavior above 50~K, which is in accordance with the Curie-Weiss law with an additional temperature independent background $\chi_0$,
\begin{equation}
  \chi = \frac{C}{T - \Theta} + \chi_0,
\end{equation}
where $C$ is the Curie constant, and is related to the effective moment $m_{\rm eff}$ as $C = Nm_{\rm eff}^2/k_{\rm B}$ with $k_{\rm B}$ being the Boltzmann constant.
The effective magnetic moment and paramagnetic Curie temperature were estimated as $m_{\rm eff} = 9.66(4)~\mu_{\rm B}$/Tb and $\Theta = 12.4(3)$~K, respectively, where $\mu_{\rm B}$ is the Bohr magneton.
The value of $m_{\rm eff}$ is very close to that of the Tb$^{3+}$ free ion ($9.72~\mu_{\rm B}$/Tb).
The positive $\Theta$ value indicates the presence of ferromagnetic interaction between the Tb$^{3+}$ magnetic moments.
At low temperatures, the magnetic susceptibility exhibits a clear anomaly at $T_{\rm C} = 11.6$~K.
The increasing FC susceptibility with decreasing temperature, together with the decreasing behavior of the ZFC susceptibility, indicates formation of the ferromagnetic (or ferrimagnetic) phase with a finite anisotropy barrier below $T_{\rm C}$.

Figure~\ref{figure3}(b) shows the magnetization curve measured at $T = 2$~K ($< T_{\rm C}$).
The inset shows the magnified view in the low-field region.
A hysteresis is clearly observed in the low field region, indicating an occurrence of the ferromagnetic/ferrimagnetic order below $T_{\rm C}$.
The coercivity is about 200~Oe, which suggests that the anisotropy barrier that pins the magnetic domain wall is quite small.
Accordingly, the remanent magnetization was found to be quite small as $0.59\mu_{\rm B}$/Tb, indicating the soft magnetic nature of Au$_{70}$Si$_{17}$Tb$_{13}$.
In the high field region, even at highest field of $H = 50000$~Oe, the magnetization does not saturate to the full moment per Tb$^{3+}$, $g_{J}\mu_{\rm B}J = 9\mu_{\rm B}$, where $J = 6$ is the total angular momentum and $g_{J} = 3/2$ is the Land\'{e} $g$ factor for the $J = 6$ states of Tb$^{3+}$.
This definitely indicates the existence of the CEF anisotropy, and is in clear contrast to the Au-Si-Gd approximant, where the CEF anisotropy is not expected due to the half filled nature of $4f$ level~\cite{TamuraR12,HirotoT14}.

Figure~\ref{figure4} shows the temperature dependence of the neutron diffraction intensity ($I_{510}$) measured at the 510 Bragg-reflection position.
At higher temperatures, the intensity does not show temperature dependence, whereas the intensity increases significantly with decreasing temperature below $T_{\rm C} \simeq 11$~K, which reasonably corresponds to the anomaly temperature in the susceptibility data [Fig.~\ref{figure3}(a)].
Note that the 510 reflection is crystallographically allowed reflection, and hence, the increasing magnetic component below $T_{\rm C}$ corresponds to the ferromagnetic order.
The inset shows the $\omega$-scan profile around the 510 peak measured at $T = 15$ and 5~K.
A significant enhancement of the Bragg peak intensity is observed below $T_{\rm C}$, superimposed on a small nuclear Bragg peak component remaining above $T_{\rm C}$.

For a number of reciprocal lattice points, integrated intensity was measured at two temperatures $T = 5$ and 15~K (paramagnetic).
Magnetic component was obtained from the difference between the two temperatures, and was converted to the squared magnetic structure factor $|F_{\rm obs}|^2$.
Figure~\ref{figure5} shows the distribution of $|F_{\rm obs}|$ in the $hk0$ plane.
Finite magnetic intensity was observed only at $h + k + l = $ even positions, whereas no magnetic intensity was detected for $h + k + l =$ odd positions.
Absence of the magnetic intensity at some half-integer and other incommensurate positions were also confirmed.
Since no extra magnetic peaks were observed other than $h + k + l =$ even positions, we conclude that the magnetic structure obeys the bcc translational symmetry.
This requires magnetic-moment configuration in all the icosahedral clusters to be identical, being consistent with the occurrence of the bulk ferromagnetic moment.

\subsection{Magnetic structure determination}

\begin{turnpage}
\begin{table*}
  \caption{\label{table2} List of BVs of all the irreducible representations for the Tb site in the Au$_{70}$Si$_{17}$Tb$_{13}$ approximant with the space group $Im\bar{3}$ and the magnetic modulation vector $q_{\rm m} = (0,0,0)$. 
    $\epsilon = (1 + \sqrt{3} i)/2$.
    Site indices for Tb atoms are defined as the $d = 1$ atom at $(0, y, z)$, 2 at $(z, 0, y)$, 3 at $(y, z, 0)$, 4 at $(-y, -z, 0)$, 5 at $(z, 0, -y)$, 6 at $(-y, z, 0)$, 7 at $(-z, 0, y)$, 8 at $(-z, 0, -y)$, 9 at $(y, -z, 0)$, 10 at $(0, -y, z)$, 11 at $(0, -y, -z)$, and 12 at $(0, y, -z)$, where $y = 0.187$ and $z = 0.306$.
  }
  \begin{ruledtabular}
    \begin{tabular}{l c c c c c c c c c c c c c}
      IR$\nu$:$\lambda$(:$\mu$) & $d = 1$ & $d = 2$ & $d = 3$ & $d = 4$ & $d = 5$ & $d = 6$ & $d = 7$ & $d = 8$ & $d = 9$ & $d = 10$ & $d = 11$ & $d = 12$ & dim \\
      \hline
IR1:1:1 & 1 0 0  & 0 1 0  & 0 0 1  & 0 0 1  & 0 -1 0  & 0 0 -1  & 0 -1 0  & 0 1 0  & 0 0 -1  & -1 0 0  & 1 0 0  & -1 0 0 & 1 \\
IR2:1:1 & 0 1 0  & 0 0 1  & 1 0 0  & -1 0 0  & 0 0 -1  & -1 0 0  & 0 0 1  & 0 0 -1  & 1 0 0  & 0 -1 0  & 0 -1 0  & 0 1 0 & 1\\
IR2:2:1 & 0 0 1  & 1 0 0  & 0 1 0  & 0 -1 0  & 1 0 0  & 0 1 0  & -1 0 0  & -1 0 0  & 0 -1 0  & 0 0 1  & 0 0 -1  & 0 0 -1 & 1\\
IR3:1:1 & 1 0 0  & 0 $-\epsilon$ 0  & 0 0 $-\epsilon^{*}$  & 0 0 $-\epsilon^{*}$  & 0 $\epsilon$ 0  & 0 0 $\epsilon^{*}$  & 0 $\epsilon$ 0  & 0 $-\epsilon$ 0  & 0 0 $\epsilon^{*}$  & -1 0 0  & 1 0 0  & -1 0 0 & 1\\
IR4:1:1 & 0 1 0  & 0 0 $-\epsilon$  & $-\epsilon^{*}$ 0 0  & $\epsilon^{*}$ 0 0  & 0 0 $\epsilon$  & $\epsilon^{*}$ 0 0  & 0 0 $-\epsilon$  & 0 0 $\epsilon$  & $-\epsilon^{*}$ 0 0  & 0 -1 0  & 0 -1 0  & 0 1 0 & 1\\
IR4:2:1 & 0 0 1  & $-\epsilon$ 0 0  & 0 $-\epsilon^{*}$ 0  & 0 $\epsilon^{*}$ 0  & $-\epsilon$ 0 0  & 0 $-\epsilon^{*}$ 0  & $\epsilon$ 0 0  & $\epsilon$ 0 0  & 0 $\epsilon^{*}$ 0  & 0 0 1  & 0 0 -1  & 0 0 -1 & 1\\
IR5:1:1 & 1 0 0  & 0 $-\epsilon^{*}$ 0  & 0 0 $-\epsilon$  & 0 0 $-\epsilon$  & 0 $\epsilon^{*}$ 0  & 0 0 $\epsilon$  & 0 $\epsilon^{*}$ 0  & 0 $-\epsilon^{*}$ 0  & 0 0 $\epsilon$  & -1 0 0  & 1 0 0  & -1 0 0 & 1\\
IR6:1:1 & 0 1 0  & 0 0 $-\epsilon^{*}$  & $-\epsilon$ 0 0  & $\epsilon$ 0 0  & 0 0 $\epsilon^{*}$  & $\epsilon$ 0 0  & 0 0 $-\epsilon^{*}$  & 0 0 $\epsilon^{*}$  & $-\epsilon$ 0 0  & 0 -1 0  & 0 -1 0  & 0 1 0 & 1\\
IR6:2:1 & 0 0 1  & $-\epsilon^{*}$ 0 0  & 0 $-\epsilon$ 0  & 0 $\epsilon$ 0  & $-\epsilon^{*}$ 0 0  & 0 $-\epsilon$ 0  & $\epsilon^{*}$ 0 0  & $\epsilon^{*}$ 0 0  & 0 $\epsilon$ 0  & 0 0 1  & 0 0 -1  & 0 0 -1 & 1\\
IR7:1:1 & 0 1 0  & 0 0 0  & 0 0 0  & 0 0 0  & 0 0 0  & 0 0 0  & 0 0 0  & 0 0 0  & 0 0 0  & 0 1 0  & 0 1 0  & 0 1 0 & 3\\
IR7:1:2 & 0 0 0  & 0 0 1  & 0 0 0  & 0 0 0  & 0 0 1  & 0 0 0  & 0 0 1  & 0 0 1  & 0 0 0  & 0 0 0  & 0 0 0  & 0 0 0 & \\
IR7:1:3 & 0 0 0  & 0 0 0  & 1 0 0  & 1 0 0  & 0 0 0  & 1 0 0  & 0 0 0  & 0 0 0  & 1 0 0  & 0 0 0  & 0 0 0  & 0 0 0 & \\
IR7:2:1 & 0 0 1  & 0 0 0  & 0 0 0  & 0 0 0  & 0 0 0  & 0 0 0  & 0 0 0  & 0 0 0  & 0 0 0  & 0 0 -1  & 0 0 1  & 0 0 -1 & 3\\
IR7:2:2 & 0 0 0  & 1 0 0  & 0 0 0  & 0 0 0  & -1 0 0  & 0 0 0  & -1 0 0  & 1 0 0  & 0 0 0  & 0 0 0  & 0 0 0  & 0 0 0 & \\
IR7:2:3 & 0 0 0  & 0 0 0  & 0 1 0  & 0 1 0  & 0 0 0  & 0 -1 0  & 0 0 0  & 0 0 0  & 0 -1 0  & 0 0 0  & 0 0 0  & 0 0 0 & \\
IR7:3:1 & 0 0 0  & 0 1 0  & 0 0 0  & 0 0 0  & 0 1 0  & 0 0 0  & 0 1 0  & 0 1 0  & 0 0 0  & 0 0 0  & 0 0 0  & 0 0 0  & 3\\
IR7:3:2 & 0 0 0  & 0 0 0  & 0 0 1  & 0 0 1  & 0 0 0  & 0 0 1  & 0 0 0  & 0 0 0  & 0 0 1  & 0 0 0  & 0 0 0  & 0 0 0 & \\
IR7:3:3 & 1 0 0  & 0 0 0  & 0 0 0  & 0 0 0  & 0 0 0  & 0 0 0  & 0 0 0  & 0 0 0  & 0 0 0  & 1 0 0  & 1 0 0  & 1 0 0 & \\
IR7:4:1 & 0 0 0  & 0 0 0  & 1 0 0  & 1 0 0  & 0 0 0  & -1 0 0  & 0 0 0  & 0 0 0  & -1 0 0  & 0 0 0  & 0 0 0  & 0 0 0 & 3\\
IR7:4:2 & 0 1 0  & 0 0 0  & 0 0 0  & 0 0 0  & 0 0 0  & 0 0 0  & 0 0 0  & 0 0 0  & 0 0 0  & 0 -1 0  & 0 1 0  & 0 -1 0 & \\
IR7:4:3 & 0 0 0  & 0 0 1  & 0 0 0  & 0 0 0  & 0 0 -1  & 0 0 0  & 0 0 -1  & 0 0 1  & 0 0 0  & 0 0 0  & 0 0 0  & 0 0 0 & \\
IR7:5:1 & 0 0 0  & 0 0 0  & 0 1 0  & 0 1 0  & 0 0 0  & 0 1 0  & 0 0 0  & 0 0 0  & 0 1 0  & 0 0 0  & 0 0 0  & 0 0 0 & 3\\
IR7:5:2 & 0 0 1  & 0 0 0  & 0 0 0  & 0 0 0  & 0 0 0  & 0 0 0  & 0 0 0  & 0 0 0  & 0 0 0  & 0 0 1  & 0 0 1  & 0 0 1 & \\
IR7:5:3 & 0 0 0  & 1 0 0  & 0 0 0  & 0 0 0  & 1 0 0  & 0 0 0  & 1 0 0  & 1 0 0  & 0 0 0  & 0 0 0  & 0 0 0  & 0 0 0 & \\
IR8:1:1 & 1 0 0  & 0 0 0  & 0 0 0  & 0 0 0  & 0 0 0  & 0 0 0  & 0 0 0  & 0 0 0  & 0 0 0  & 1 0 0  & -1 0 0  & -1 0 0 & 3\\
IR8:1:2 & 0 0 0  & 0 1 0  & 0 0 0  & 0 0 0  & 0 1 0  & 0 0 0  & 0 -1 0  & 0 -1 0  & 0 0 0  & 0 0 0  & 0 0 0  & 0 0 0 & \\
IR8:1:3 & 0 0 0  & 0 0 0  & 0 0 1  & 0 0 -1  & 0 0 0  & 0 0 1  & 0 0 0  & 0 0 0  & 0 0 -1  & 0 0 0  & 0 0 0  & 0 0 0 & \\
IR8:2:1 & 0 0 0  & 1 0 0  & 0 0 0  & 0 0 0  & -1 0 0  & 0 0 0  & 1 0 0  & -1 0 0  & 0 0 0  & 0 0 0  & 0 0 0  & 0 0 0 & 3\\
IR8:2:2 & 0 0 0  & 0 0 0  & 0 1 0  & 0 -1 0  & 0 0 0  & 0 -1 0  & 0 0 0  & 0 0 0  & 0 1 0  & 0 0 0  & 0 0 0  & 0 0 0 & \\
IR8:2:3 & 0 0 1  & 0 0 0  & 0 0 0  & 0 0 0  & 0 0 0  & 0 0 0  & 0 0 0  & 0 0 0  & 0 0 0  & 0 0 -1  & 0 0 -1  & 0 0 1 & \\
IR8:3:1 & 0 0 0  & 0 0 1  & 0 0 0  & 0 0 0  & 0 0 1  & 0 0 0  & 0 0 -1  & 0 0 -1  & 0 0 0  & 0 0 0  & 0 0 0  & 0 0 0 & 3\\
IR8:3:2 & 0 0 0  & 0 0 0  & 1 0 0  & -1 0 0  & 0 0 0  & 1 0 0  & 0 0 0  & 0 0 0  & -1 0 0  & 0 0 0  & 0 0 0  & 0 0 0 & \\
IR8:3:3 & 0 1 0  & 0 0 0  & 0 0 0  & 0 0 0  & 0 0 0  & 0 0 0  & 0 0 0  & 0 0 0  & 0 0 0  & 0 1 0  & 0 -1 0  & 0 -1 0 & \\
IR8:4:1 & 0 0 0  & 0 0 0  & 0 0 1  & 0 0 -1  & 0 0 0  & 0 0 -1  & 0 0 0  & 0 0 0  & 0 0 1  & 0 0 0  & 0 0 0  & 0 0 0 & 3\\
IR8:4:2 & 1 0 0  & 0 0 0  & 0 0 0  & 0 0 0  & 0 0 0  & 0 0 0  & 0 0 0  & 0 0 0  & 0 0 0  & -1 0 0  & -1 0 0  & 1 0 0 & \\
IR8:4:3 & 0 0 0  & 0 1 0  & 0 0 0  & 0 0 0  & 0 -1 0  & 0 0 0  & 0 1 0  & 0 -1 0  & 0 0 0  & 0 0 0  & 0 0 0  & 0 0 0 & \\
    \end{tabular}
  \end{ruledtabular}
\end{table*}
\end{turnpage}

\begin{figure}
  \includegraphics[scale=0.3, angle=-90]{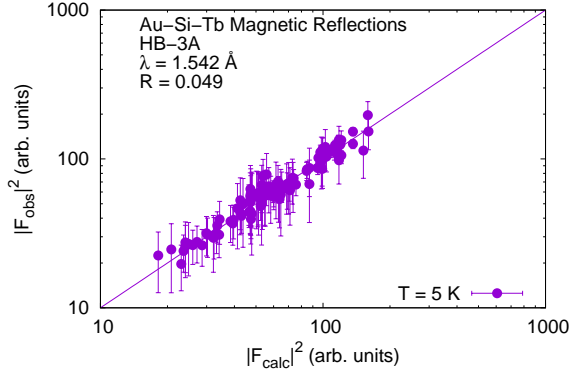}
  \caption{\label{figure6} $|F_{\rm obs}|^2$ versus $|F_{\rm calc}|^2$ for magnetic Bragg reflections.
    The magnetic $|F_{\rm obs}|^2$ were obtained by subtracting paramagnetic ($T = 15$~K) intensity from the low-temperature ($T = 5$~K) data.
    The calculated $|F_{\rm calc}|^2$ were obtained using the parameters listed in Table~\ref{table3}.
    The conventional $R$-factor is 0.049, whereas weighted $\chi^2$ is 0.26.
  }
\end{figure}

\begin{figure*}
  \includegraphics[scale=0.7, angle=90, trim=0 100 100 100]{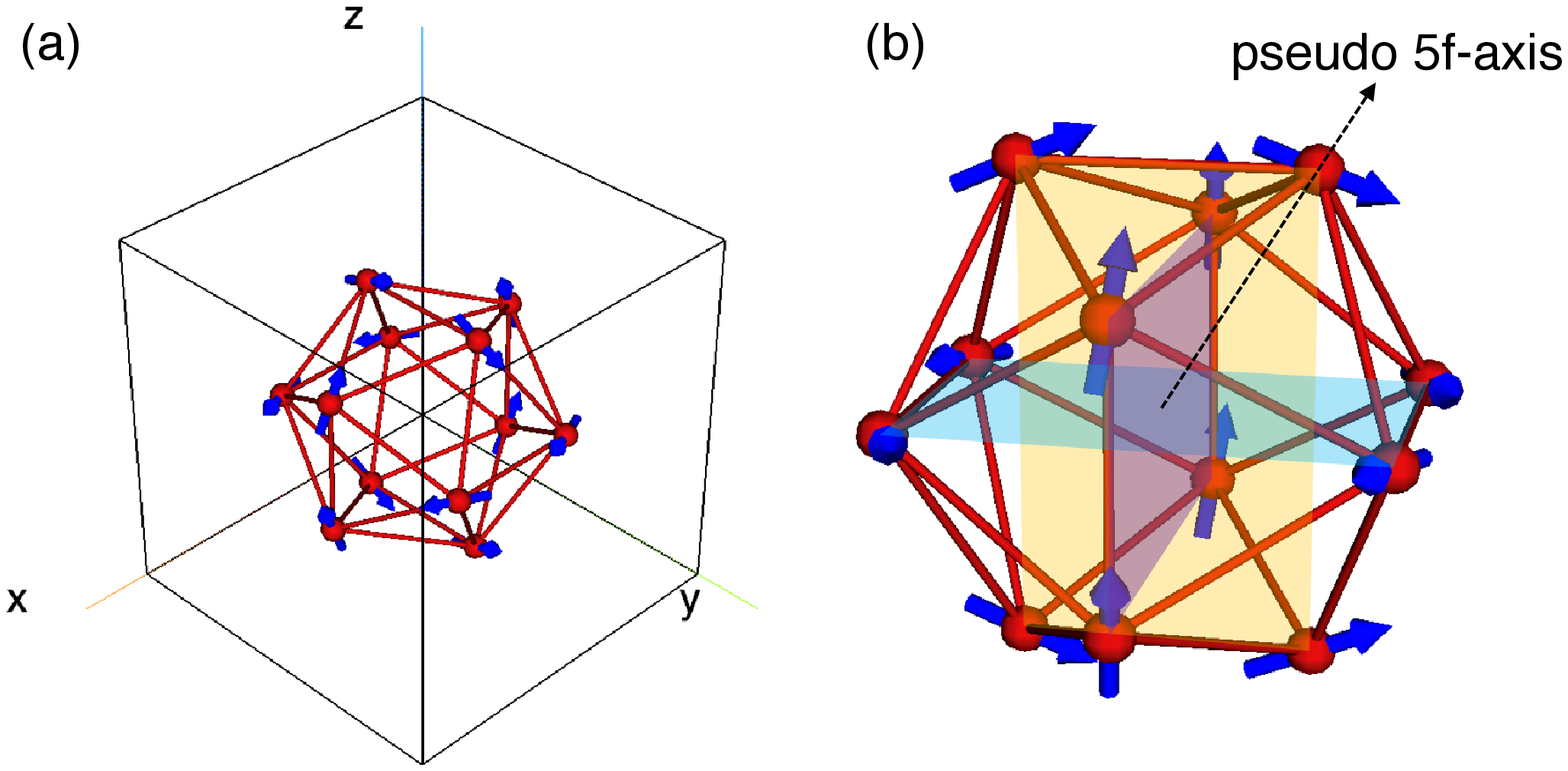}
  \caption{\label{figure7}
    (a) Magnetic structure proposed in the present study.
    The spin configuration of the domain 1 is illustrated on the icosahedral spin cluster at the body center position.
    (b) Magnified view of the single cluster with the emphasis on the three rectangular planes on which the spins have uncompensated ferromagnetic components.
    Each rectangular unit has net moment along the crystallographic $a$-, $b$- or $c$-axis, resulting in the net ferromagnetic moment of the icosahedral cluster along the $[111]$ direction for this domain.
    One of the (pseudo) $5f$-axis is depicted by the dashed lines.  
    It may be noted that all the axes from the center of the (pseudo) icosahedron to the vertices are (pseudo) $5f$-axes.
  }
\end{figure*}

\begin{table}
  \caption{\label{table3} Refined  parameters for the magnetic structure at $T = 5$~K.
    Number of magnetic Bragg reflections used in the refinement was 100 ($I > 2\sigma$), whereas number of refined parameters was 8.
    The conventional $R$-factor is $R = 0.049$, whereas weighted $\chi^2 = 0.26$.
  }
  \begin{ruledtabular}
    \begin{tabular}{l c c c c c c}
      Parameter & Refined value \\ \hline 
      Moment size & 6.9 ($\mu_{\rm B}$)\\
      Domain 1 population & 0.22(4) \\
      Domain 2 population & 0.30(4) \\
      Domain 3 population & 0.24(4) \\
      Domain 4 population & 0.24 \\
      Coefficient for IR7:1 & 6.67(11) ($\mu_{\rm B}$)\\
      Coefficient for IR7:2 & -2.63(34) ($\mu_{\rm B}$)\\
      Coefficient for IR7:3 & 0.84(25) ($\mu_{\rm B}$)\\
      Coefficient for IR7:4 & -0.19(68) ($\mu_{\rm B}$)\\
      Coefficient for IR7:5 & 0.44(19) ($\mu_{\rm B}$)\\
    \end{tabular}
  \end{ruledtabular}
\end{table}

The initial magnetic structure model was obtained by using the magnetic representation analysis~\cite{Izyumov1979,Izyumov1991}.
In the representation analysis, one assumes that the magnetic structure may be modeled by a linear combination of magnetic basis vectors (BVs) belonging to a single irreducible representation of the ``$\vec{k}$-group'' of underlying crystallographic space group.
Since the magnetic unit cell is the same as the chemical unit cell, conserving the bcc centering-translational symmetry, the magnetic modulation vector should be $\vec{q}_{\rm m} = (0, 0, 0)$.
For this magnetic modulation vector, the $\vec{k}$-group coincides with the crystallographic space group itself ($Im\bar{3}$), and magnetic representations is reduced into six one-dimensional representations and two three-dimensional representations.
The BVs for the irreducible magnetic representations are listed in Table~\ref{table2}.

Using the BVs, the initial magnetic structure may be modeled as:
\begin{equation}\label{eq:magneticstructure1}
  \langle\vec{J}_{\vec{l},\vec{\tau}, \vec{d}}\rangle = \frac{J}{2} \left \{
    \vec{a}_{\vec{d}} \exp \left [- {\rm i} \vec{q}_{\rm m} 
      \cdot (\vec{l} + \vec{\tau}) \right] 
    + {\rm c.c.}\right \},
\end{equation}
where $\langle\vec{J}_{\vec{l}, \vec{\tau}, \vec{d}}\rangle$ stands for the ordered spin direction (vector) at the $\vec{d} + \vec{\tau}$ site in the unit cell at the position $\vec{l}$, and $\vec{\tau} = (0,0,0)$ or $(1/2, 1/2, 1/2)$, which are the centering translation vectors for the bcc lattice.
The complex magnetic moment vector $\vec{a}_{\vec{d}}$ is given as a linear combination of all the BVs $\vec{\Psi}^{\vec{d}}_{\nu, \lambda (\mu)}$ in a given ($\nu$-th) irreducible representation as:
\begin{equation}\label{eq:magneticstructure2}
  \vec{a}_{\vec{d}} = \sum_{\lambda (\mu)} C_{\lambda (\mu)}^{\nu} \vec{\Psi}^{\vec{d}}_{\nu, \lambda (\mu)},
\end{equation}
where $C_{\lambda (\mu)}^{\nu}$ is the linear combination coefficient for a given BV, $\lambda$ numbers each BV,  and $\mu (= 1,2,3)$ specifies its component only used for the three dimensional BVs ($\nu = 7$ and 8).
From the table, it can be immediately seen that only the seventh irreducible representation ($\nu = 7$ or IR7) can be compatible with the the bulk ferromagnetic moment; all the other representations give rise to compensating antiferromagnetic structures.
Hence, we only test a magnetic structure model consisting of the BVs in the $\nu = 7$ irreducible representation (i.e., $C^{7}_{\lambda (\mu)}$ only in Eq.~\ref{eq:magneticstructure2}) for the magnetic structure analysis.
We further assumed that the ordered moment sizes of all the Tb$^{3+}$ ions are identical; this was most simply achieved by setting $C^{7}_{\lambda (1)} = C^{7}_{\lambda (2)}  = C^{7}_{\lambda (3)}$.
Consequently, five independent coefficients ($C^{7}_{\lambda (1)}$ with $\lambda = 1, ..., 5$) were used as adjustable parameters.
It should be noted that the magnetic structure given by Eqs.~\ref{eq:magneticstructure1} and \ref{eq:magneticstructure2} breaks the original crystallographic (paramagnetic) symmetry;
Hence, independent magnetic structures obtained by operating the crystallographic (paramagnetic) symmetry operations form magnetic domains.
Under the assumption of equal moment size, all the 48 symmetry operations of $Im\bar{3}$ results in eight independent magnetic domains associated with the bulk magnetic-moment directions $[111], [\bar{1}11], [1\bar{1}1] [11\bar{1}], [\bar{1}\bar{1}1], [\bar{1}1\bar{1}], [1\bar{1}\bar{1}]$ and $[\bar{1}\bar{1}\bar{1}]$.
Among them, those related by the time-reversal symmetry cannot be distinguished by neutron diffraction, and hence we introduce four parameters, $v_n$, for the domain volume fraction, of which only three are independent due to the condition $\sum_n v_n = 1$ for $n = 1, ..., 4$.
All the other parameters, including isotropic atomic displacement and secondary extinction parameters, were fixed to the values obtained in the crystal structure refinement.

The least-square fitting was performed to $|F_{\rm obs}|^2$ using the above model structure with the coefficient $C^{7}_{\lambda (\mu)}$ and domain volume fractions $v_n$ as adjustable parameters.
The resulting calculated optimal $|F_{\rm calc}|^2$ is compared with $|F_{\rm obs}|^2$ in Fig.~\ref{figure6}.
The satisfactorily linear correspondence can be seen in the observed and calculated structure factors.
The conventional $R$-factor was 0.049, which is almost the same value as that obtained for the crystal structure refinement shown in Fig.~\ref{figure2}.
The resulting refined magnetic-structure parameters are given in Table~\ref{table3}.
The ordered moment size is rather small as $6.9\mu_{\rm B}$ compared to the free Tb$^{3+}$ moment size $g_{J}\mu_{\rm B}J = 9\mu_{\rm B}$.
This may be partly due to the relatively high temperature ($5$~K) at which the magnetic intensity was collected, and also partly due to the CEF effect discussed later.
It can be also seen from the refined parameters that the volume fractions of the four domains are mostly the same.
The dominant component of the ordered moment is given by the first BV of IR7, $\vec{\Psi}_{7, 1}$, to which small contribution from the second BV $\vec{\Psi}_{7,2}$ is added.
Other contributions from the other three BVs are relatively small, and thus may be ignored.
It may be noted that $\vec{\Psi}_{7,1}$ gives rise to the bulk ferromagnetic moment, whereas $\vec{\Psi}_{7,2}$ corresponds to strictly antiferromagnetic order, and hence cannot give rise to the bulk ferromagnetic moment.
Both $\vec{\Psi}_{7,1}$ and $\vec{\Psi}_{7,2}$ have their ordered magnetic moment being in the local mirror plane at the Tb$^{3+}$ sites.
Hence, we can conclude that the ordered magnetic moments are dominantly in the local mirror plane.

\begin{figure*}
\includegraphics[scale=0.34, angle=-90, trim={0cm 0cm 0cm 0cm}]{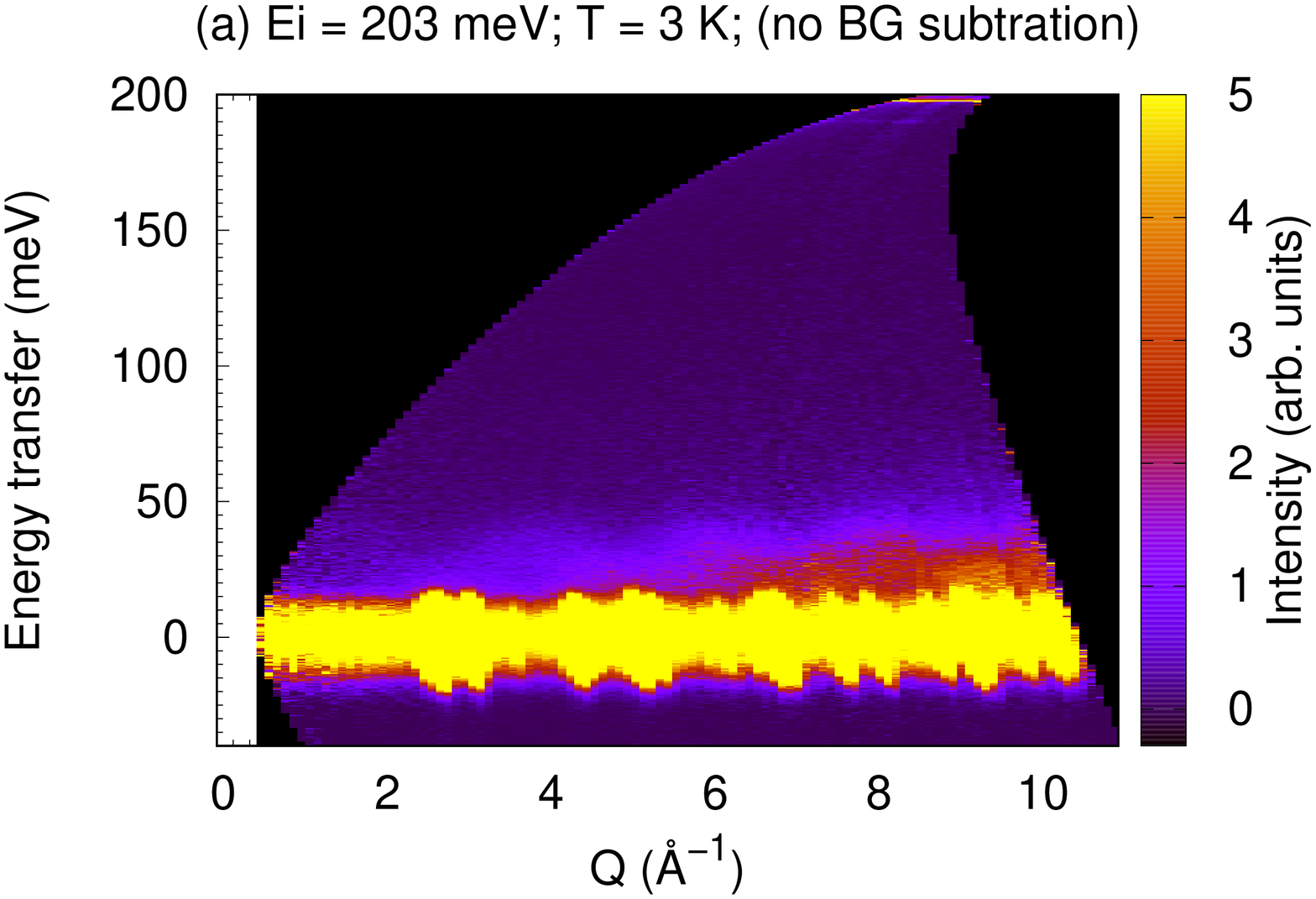}
\includegraphics[scale=0.34, angle=-90, trim={0cm 0cm 0cm 0cm}]{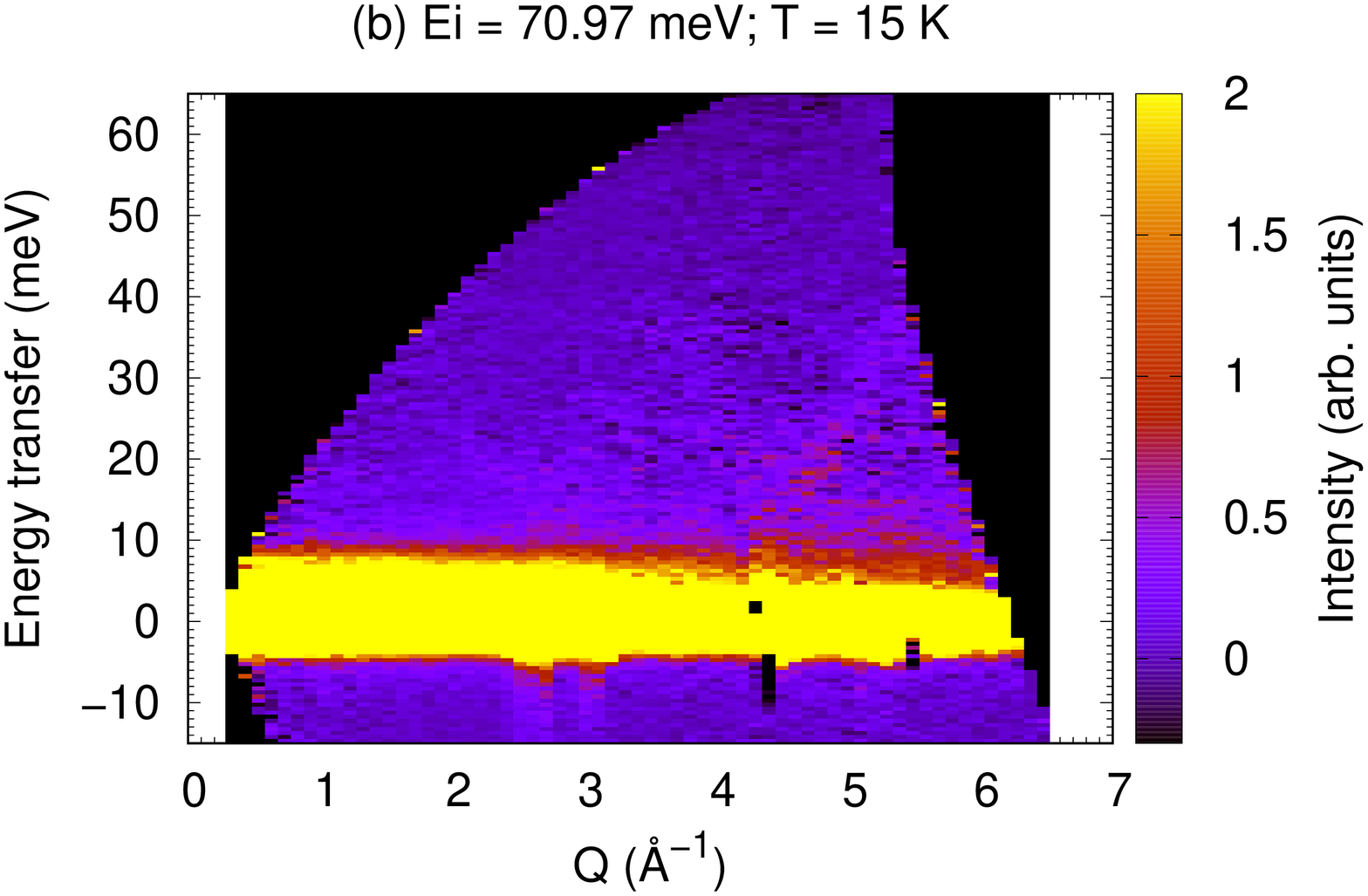}
\includegraphics[scale=0.34, angle=-90, trim={0cm 0cm 0cm 0cm}]{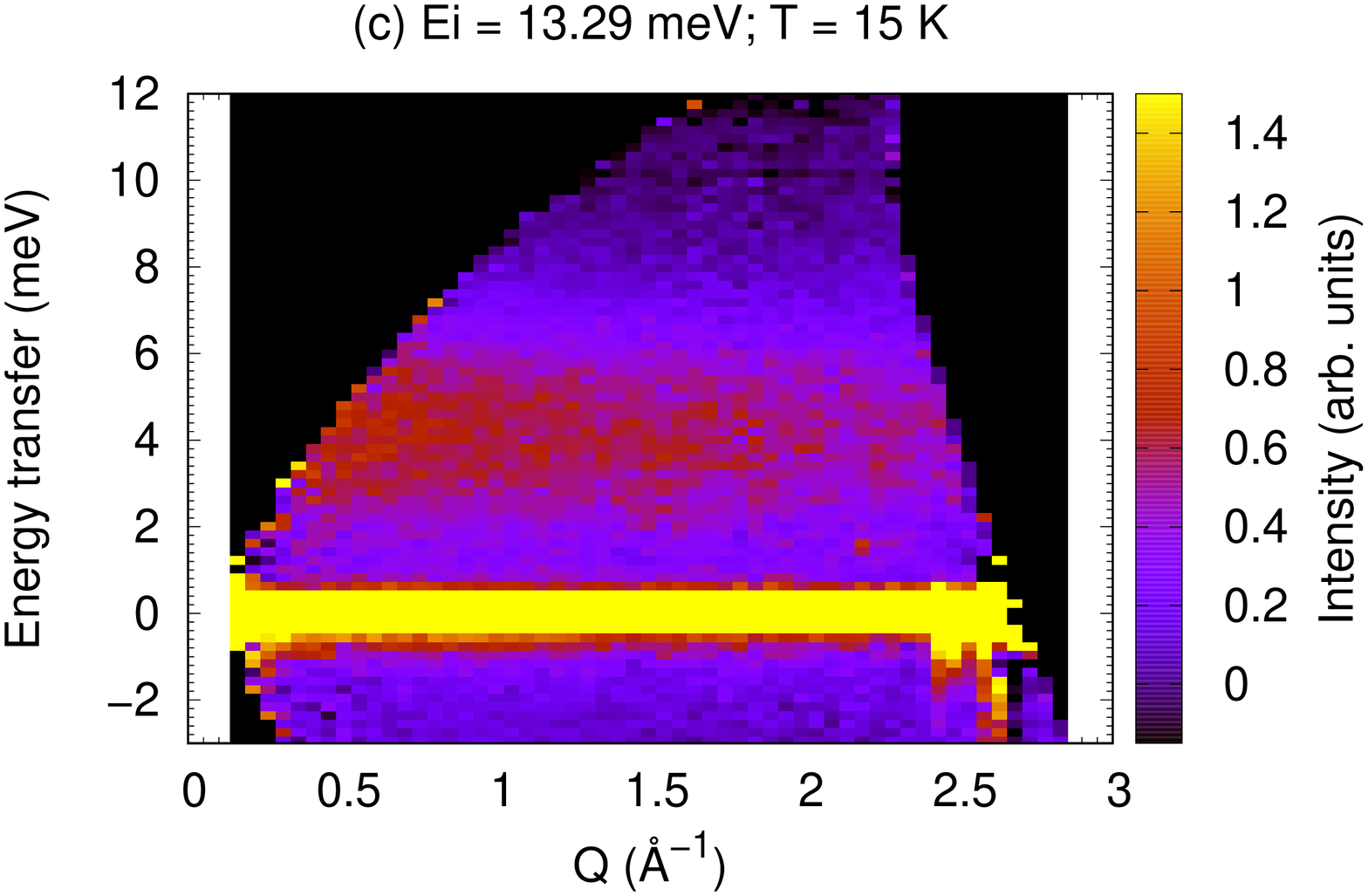}
\includegraphics[scale=0.34, angle=-90, trim={0cm 0cm 0cm 0cm}]{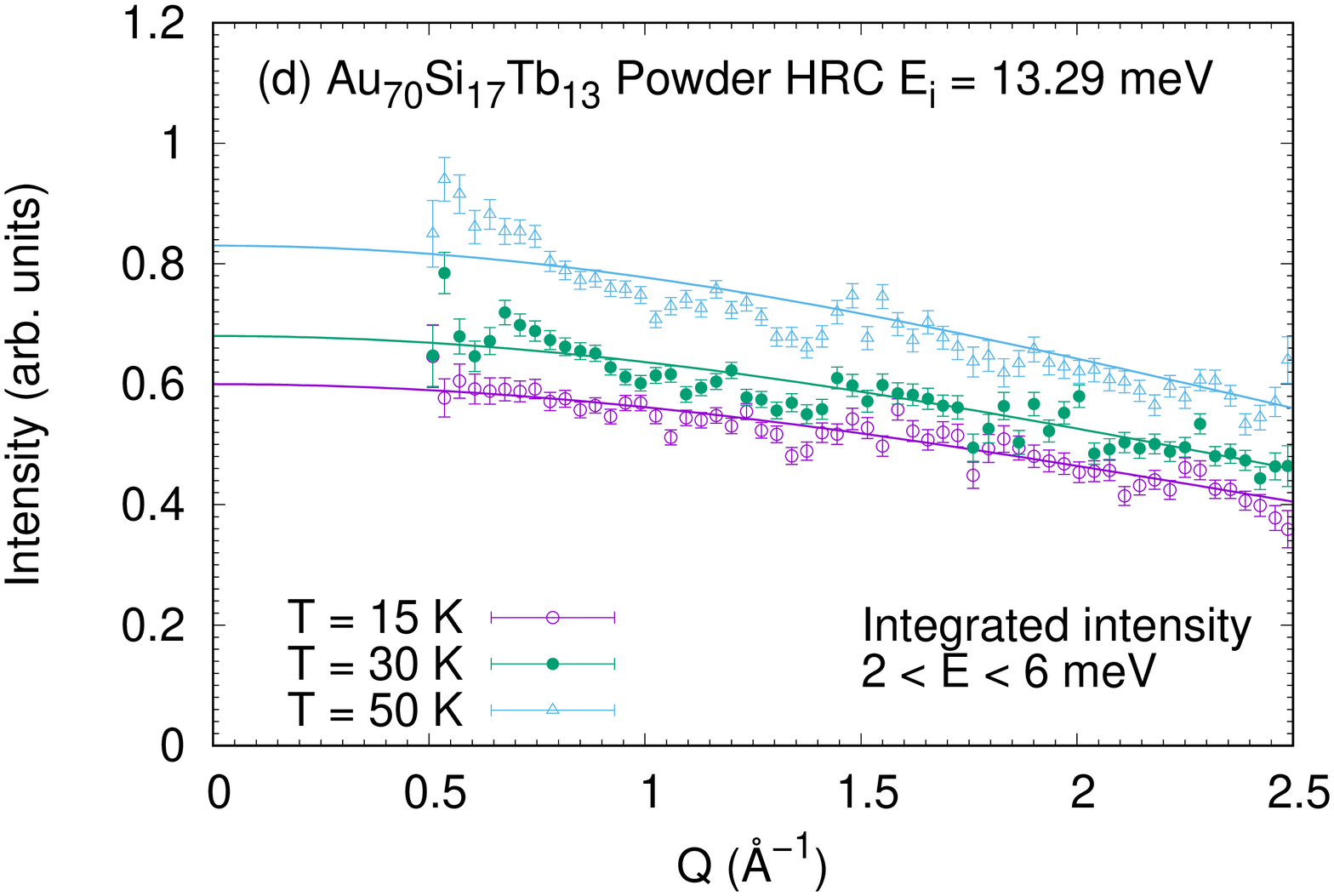}
\caption{(Color online) (a-c) Neutron inelastic scattering spectra observed with the three different incident energies $E_{\rm i} = 13.29$, 70.97, and 203~meV.
  The measurement temperatures were 3~K for (a), whereas 15~K for (b) and (c).
  The background was estimated by measuring the empty Al-can, and was subtracted from the data shown in (b) and (c), whereas for (a) background subtraction was not performed.
  (d) $Q$-dependence of the inelastic intensity integrated in a range $2 < \hbar \omega < 6$~meV measured at $T = 15$~K, 30~K and 50~K.
  The solid lines stand for the square of the magnetic form factor for the Tb$^{3+}$ ion scaled to the observations.
}\label{figure8}
\end{figure*}

The refined magnetic structure in one icosahedral cluster at the body-center position is illustrated in Fig.~\ref{figure7}.
As noted above, the dominant components of the ordered moments are in the local mirror planes, which is depicted by the semi-transparent rectangular planes in the figure.
Ordered moment direction is tilted from the crystallographic axis (either $\vec{a}, \vec{b}$, or $\vec{c}$, depending on the Tb$^{3+}$ site) due to the finite $\vec{\Psi}_{7,2}$ component, and tilting angle is estimated as $\sim 22^{\circ}$ from the corresponding (nearest) crystallographic axis.
Or, if we measure the angle between the moment direction and the pseudo $5f$-axis, which is the axis passing through the origin and the Tb$^{3+}$ site, then the moment direction is approximately 80 degrees away from the axis.
Hence, we can conclude that the moment direction is nearly perpendicular to the local pseudo $5f$-axis.
Apparently, the refined magnetic structure is non-collinear and non-coplanar, as found in the antiferromagnetic Au$_{72}$Al$_{14}$Tb$_{14}$ approximant, and cannot be a simple collinear ferrimagnetic structure envisaged earlier.

Detailed comparison of the presently refined magnetic structure to that obtained for Au$_{72}$Al$_{14}$Tb$_{14}$ approximant may be informative.
First, the ordered moment direction presently determined in the ferrimagnetic Au$_{70}$Si$_{17}$Tb$_{13}$ approximant is mostly parallel or antiparallel to that observed in the antiferromagnetic Au$_{72}$Al$_{14}$Tb$_{14}$ approximant; only approximately 5 degrees (or 175 degrees when antiparallel) difference in the ordered moment directions in the two compounds.
Hence, we can speculate that the local anisotropy direction, possibly due to the local CEF, is mostly the same.
Secondly, on the other hand, there are apparent difference in the ordered moment arrangement in a single icosahedral cluster; the magnetic moments at the opposite vertices are parallel (ferromagnetic) in the present Au$_{70}$Si$_{17}$Tb$_{13}$, in striking contrast to the antiparallel (antiferromagnetic) arrangement found in the Au$_{72}$Al$_{14}$Tb$_{14}$ approximant.
This suggests delicate balance of the exchange coupling results in the decisive difference in the bulk magnetic properties.
We further note that the non-coplanar ferrimagnetic order in the present Au$_{70}$Si$_{17}$Tb$_{13}$ does not break the inversion symmetry around the cluster center, which is also in striking contrast to that in the Au$_{72}$Al$_{14}$Tb$_{14}$ approximant.

\subsection{Neutron inelastic scattering spectra}

The magnetic structure analysis indicates that the ordered moment of Tb$^{3+}$ is in the local mirror plane, and is nearly perpendicular to the pseudo $5f$-axis.
This suggests existence of strong easy-axis anisotropy along this direction for the Tb$^{3+}$ magnetic moment.
For the rare-earth compounds, the anisotropy usually originates from the CEF splitting of the ground $J$-multiplet for the open-shell $4f$ electrons.
Such energy splitting can be most effectively studied by neutron inelastic scattering, and hence we have performed the neutron inelastic scattering experiment on the powder sample of the Au$_{70}$Si$_{17}$Tb$_{13}$ approximant.

Neutron inelastic scattering was measured at several temperatures in the range $3 < T < 100$~K.
As the representative results, Figs.~\ref{figure8}(a-c) show the inelastic spectra observed with three different incident neutron energies $E_{\rm i} = 13.29$, $70.97$ and $200$~meV.
No significant magnetic signal can be seen in the high energy regions shown in Figs.~\ref{figure8}(a) and \ref{figure8}(b); only weak dispersive signal is seen up to $\hbar \omega = 40$~meV in the high-$Q$ region, which may be a phonon contribution.
In the low energy region shown in Fig.~\ref{figure8}(c), inelastic excitation can be seen around $\hbar\omega \simeq 4$~meV.
This is the only magnetic signal we found in the present inelastic-scattering experiment.

The $Q$-dependence of the $\hbar \omega \simeq 4$~meV peak was checked by taking an energy integration of the inelastic spectra in $2 < \hbar\omega < 6$~meV.
The results for the selected temperatures $T = 15$, 30 and 50~K are shown in Fig.~\ref{figure8}(d).
It is clear in the figure that the scattering function is almost independent of $Q$ except for the trivial Tb$^{3+}$ magnetic form factor (indicated by the solid lines)~\cite{FreemanAJ79}.
It should be noted that the $Q$ dependence is insignificant even at $T = 15$~K, which is quite close to the ferrimagnetic transition temperature $T_{\rm C}$.
This suggests that inter-site spin correlations are not significant in this Au$_{70}$Si$_{17}$Tb$_{13}$ approximant, and that the broad inelastic peak observed at $\hbar \omega = 4$~meV may likely originate from the local single-site transition.
As speculated in the beginning of this subsection, for the $4f$-electrons of Tb$^{3+}$, such local energy levels may be the CEF splitting levels.

\begin{figure}
\includegraphics[scale=0.35, angle=-90, trim={0cm 0cm 0cm 0cm}]{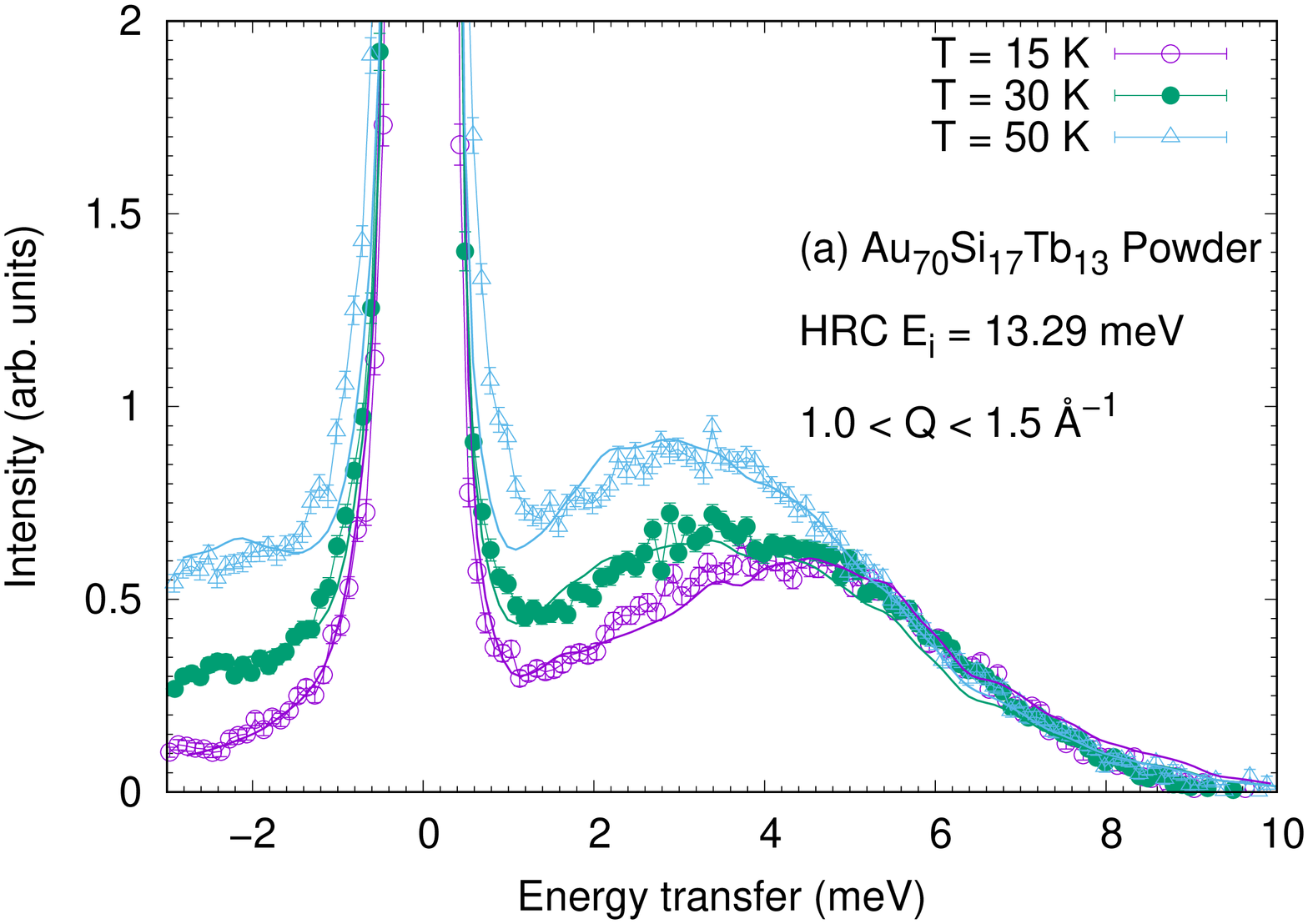}
\includegraphics[scale=0.35, angle=-90, trim={0cm 0cm 0cm 0cm}]{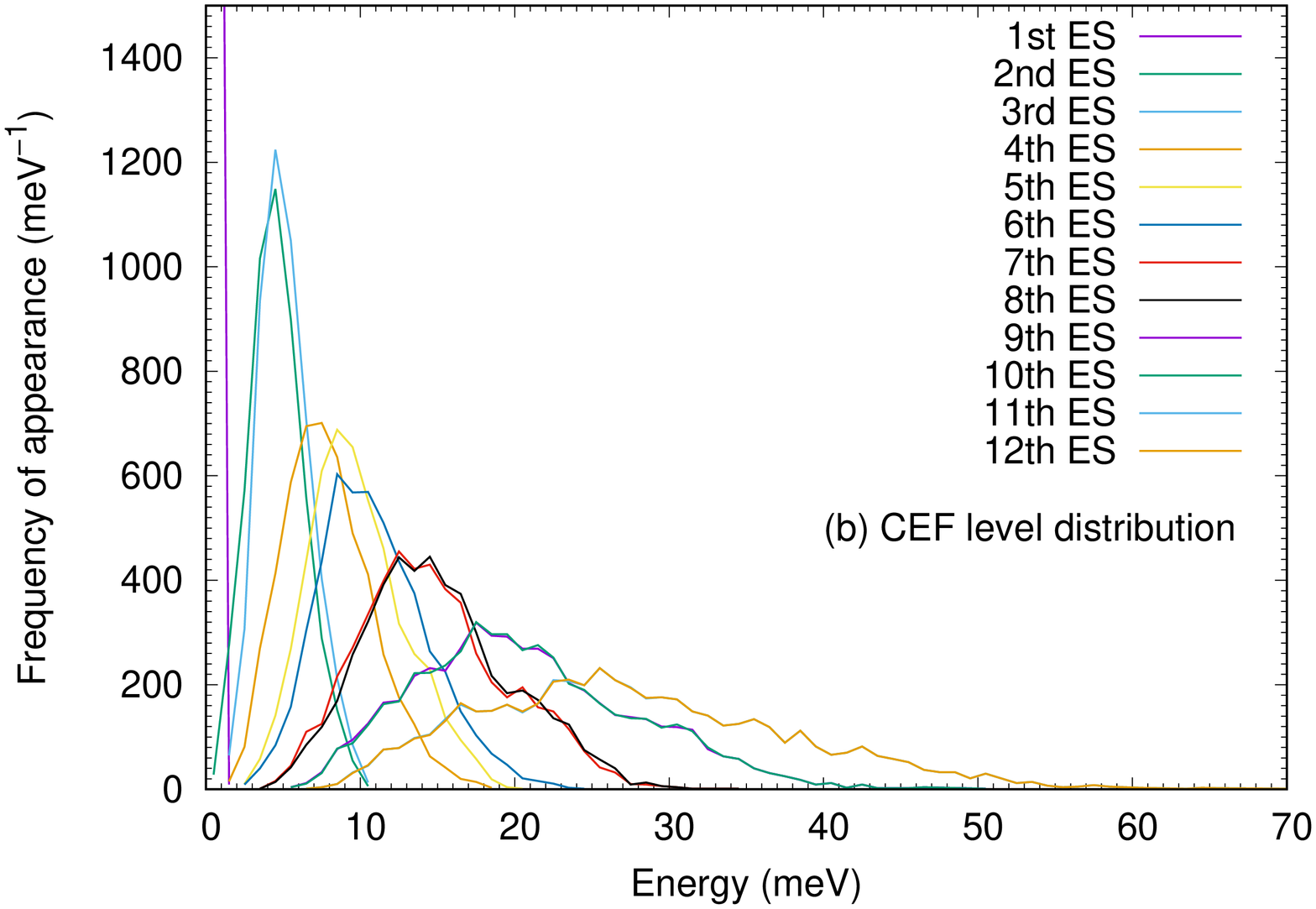}
\caption{(Color online) (a) Inelastic neutron scattering spectra at $T = 15, 30$, 50~K in the Au$_{70}$Si$_{17}$Tb$_{13}$ approximant.
  The calculated CEF excitation spectra using the point-charge model incorporating chemical disorder are platted by the solid lines for the corresponding temperatures.
  (b) Frequency of appearance of the CEF excited states (ES) as a function of energy, estimated in the present point-charge calculation.
  Most of the ESs are approximately doubly degenerated, except the 4th, 5th and 6th ESs.
  There are 5000 variations for each ES, corresponding to the 5000 different atomic configurations around the Tb$^{3+}$ ions assumed in the present point-charge calculation.
}\label{figure9}
\end{figure}

To better visualize the temperature dependence of the inelastic scattering, $Q$-integrated energy spectra were obtained with the integration range $1.0 < Q < 1.5$~\AA$^{-1}$.
The resulting energy spectra for the temperatures $T = 15$, 30, and 50~K are shown in Fig.~\ref{figure9}(a).
As already noted above, at the lowest paramagnetic temperature $T = 15$~K, there appears only a single broad peak around $\hbar\omega \simeq 4$~meV.
It should be noted that the energy resolution for this experiment is $\Delta \hbar\omega \simeq 0.4$~meV (FWHM) at the elastic position, and hence, the width of the inelastic peak cannot be explained by the instrumental resolution, but has to be regarded as intrinsic.
Indeed, the elastic peak is much sharper compared to the inelastic signal.
As the temperature is increased, the broad peak shifts to lower energies with drastic increase of the scattering intensity in the low energy region ($\hbar \omega < 4$~meV).

For local excitations, such as the CEF excitations, the broadening is often attributed to finite lifetime of the ground and excited states due to scattering by other degree of freedom, such as phonons or conduction electrons.
The present experiment shows that the broadening occurs even at the lowest temperature where phonons of a few meV are already suppressed significantly, and hence, the phonon scattering is unlikely origin of the CEF peak broadening.
The hybridization and/or spin scattering by conduction electrons may give rise to the broadened CEF peaks at very low temperatures, however, we found that the elastic peak is mostly resolution limited, indicating negligible effect of such hybridization.
These results suggest that a certain mechanism other than the shortened lifetime is in effect.
In addition, the increasing behavior of the inelastic scattering intensity at higher temperatures is unusual, since in general the inelastic scattering intensity between the CEF splitting levels decreases at high temperatures, as it depends on the population of the ground state.
Hence, although the $Q$-independence strongly indicates that the inelastic signal is of the CEF origin, above issues suggest further detailed analysis is in order to conclude the CEF origin.
In the next section, we propose a possible microscopic CEF model which explains the broadness of the inelastic peak and its intriguing temperature dependence, and will show that such a model is also consistent with the ordered moment direction determined in the previous subsection.

\subsection{Point-charge analysis for the CEF excitations}

\begin{figure}
\includegraphics[scale=0.35, angle=90, trim={1.5cm 2cm 1.5cm 3cm}]{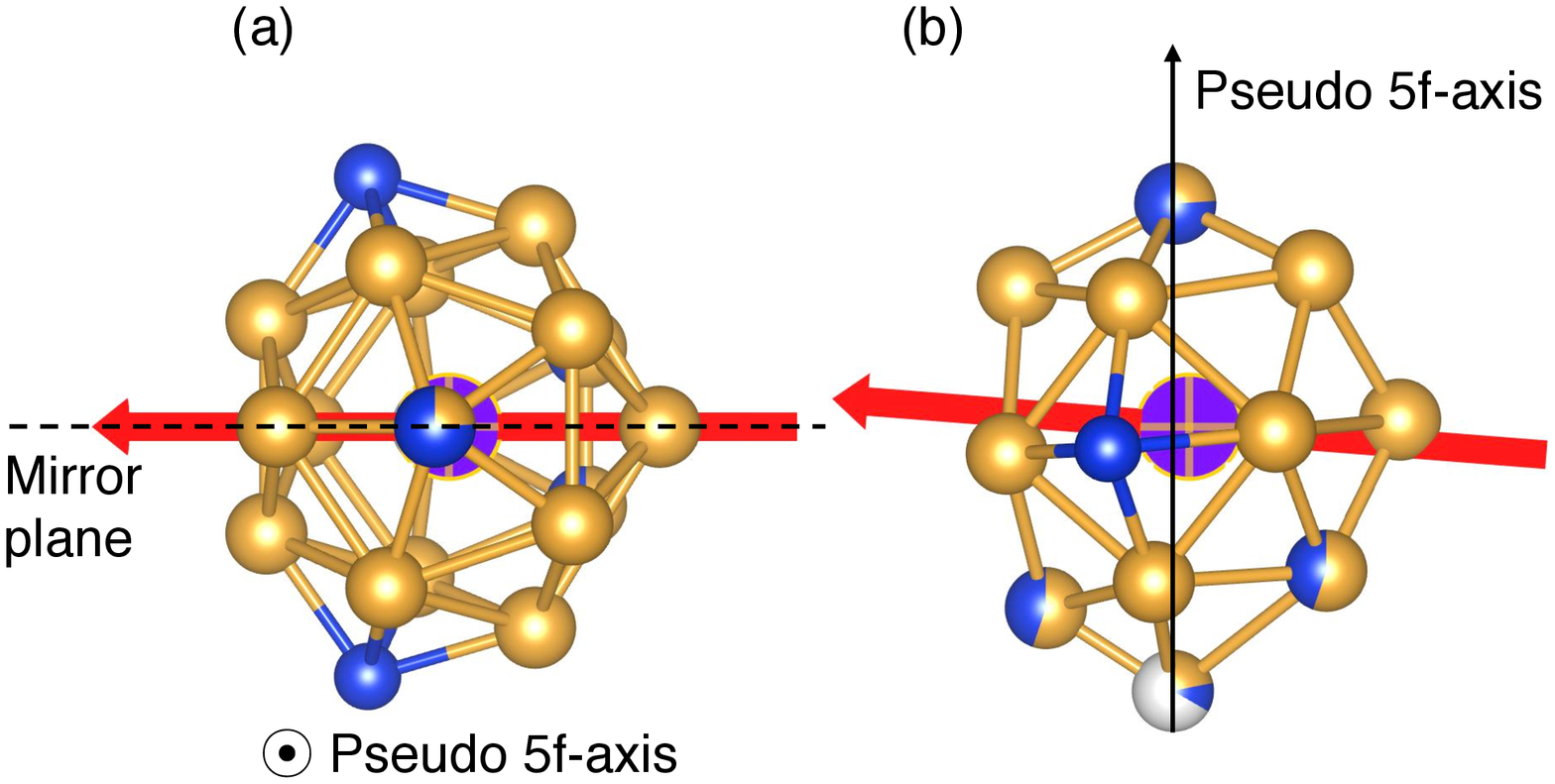}
\caption{(Color online) Atom configuration around the Tb$^{3+}$ ion.
  (a) A view along the pseudo $5f$-axis.
  (b) A view normal to the mirror plane.
  Yellow, blue and violet spheres denote Au, Si and Tb atoms, respectively.
  The fractionally occupied sites are indicated by shared colors on spheres.
  The red arrows indicate ordered moment direction determined in the magnetic structure analysis, whereas the dotted line indicate local mirror plane around the Tb$^{3+}$.
  Directions of the pseudo $5f$-axis are shown in the figure, too.
}\label{figure10}
\end{figure}

Quite often, the CEF excitations are analyzed using the operator equivalent technique~\cite{HutchingsMT64}; the CEF Hamiltonian is rewritten in terms of the Stevens' operator equivalents $\hat{O}^q_k$ with their coefficients as adjustable parameters:  ${\cal H}_{\rm CEF} = \sum_{k,q} B^q_k \hat{O}^q_k$ ($k = 2,4,6$ and $q = -k, -k+1, ..., k$).
For a rare-earth ions at a high-symmetry site, the condition that the CEF Hamiltonian has to be invariant under point group operations greatly reduces the number of non-zero coefficients, i.e., adjustable parameters.
However, in the present Au$_{70}$Si$_{17}$Tb$_{13}$ approximant, the Tb$^{3+}$ site only has a mirror symmetry $m$ ($C_s$), and hence, only one condition $B^q_{-k} = B^q_{k}$ is imposed even the quantization axis is appropriately taken.
Furthermore, there are several sites with fractional Au/Si occupancy in the vicinity of the Tb$^{3+}$ sites (see Fig.~\ref{figure10}).
This introduces local mirror-symmetry breaking, and consequently, different CEF potential for different Tb$^{3+}$ site depending on the local Au/Si arrangement at the fractional sites.
Hence, with the Stevens' operator equivalent method, we have to deal with all the $B^q_k\hat{O}^q_k$ terms, and much complicatedly,  they have to be taken as site dependent.
This method, hence, includes a huge number of adjustable parameters, and practically is impossible to be pursued.

To reduce the adjustable parameter number, as well as to incorporate with the site-dependent CEF potential, we, here, use the point-charge model for the CEF energy level calculation taking account of the statistically distributed Au and Si elements at the fractionally occupied sites.
The details of the calculation for the energy levels and corresponding $4f$ wavefunctions are given in Appendix A.

In the present CEF calculation, it is assumed that the fractional sites (Au6/Si6, Au7/Si7, and Au8/Si8 sites listed in Table~\ref{table1}) within the radius $R_{\rm in} = 6$~\AA\ from the center Tb$^{3+}$ atom are statistically occupied either by Au or Si (or left vacant) with the probability given by the occupancy.
For the fractional sites between $R_{\rm in} < R < R_{\rm max} = 30$~\AA, we use occupancy-weighted averages of Au and Si valences.
In total $n_{\rm conf} = 5000$ atomic configurations were generated.

For the $n_{\rm conf}$ atomic configurations, the CEF energy levels $E^i_{n}$ and corresponding wavefunctions $|n\rangle^i$ were calculated, where $i$ numbers each configuration.
The configuration-averaged scattering intensity for the transitions between the CEF splitting levels from a powder sample may be given as:
\begin{widetext}
\begin{equation}\label{eq:CEFSQw}
  \left ( \frac{{\rm d}^2\sigma}{{\rm d}\Omega{\rm d}E_{\rm f}}\right )_{\rm inel}
  =
  \frac{2}{3} (\gamma r_0)^2 \left [ \frac{1}{2} g_J f_{\rm mag}(Q) \right ]^2 \frac{N}{n_{\rm conf}}\sum_i \sum_{nm \alpha} \frac{\exp(-E^i_n/k_{\rm B}T)}{Z^i} |\langle m | \hat{J}^{\alpha} | n \rangle_i |^2 \delta(\hbar \omega - E^i_m + E^i_n),
\end{equation}
\end{widetext}
where $\alpha = x, y$, or $z$, $k_{\rm B}$ is the Boltzmann constant, $\gamma = -1.91$, and $r_0$ is the classical radius of the electron.
$Z^i$ is the partition function for the $i$-th configuration.
$N$ and $f_{\rm mag}(Q)$ are the number and the magnetic form factor~\cite{FreemanAJ79} of the Tb$^{3+}$ ions, respectively.
As seen above, we ignored intrinsic line widths of the CEF levels; this point will be discussed later.
The calculated $S(Q, \hbar\omega)_{\rm inel}$ is further convoluted by the instrumental resolution function estimated from the elastic peak shape.

Least square fitting to the $Q$-integrated spectra measured at $T = 15$, 30 and 50~K was performed with the point-charge parameters $q_{\rm Au}, q_{\rm Si}$ and $q_{\rm Tb}$ (for Au, Si and Tb, respectively) being adjustable parameters.
In the fitting, we assumed that the point charges for one element species are the same regardless of their site symmetry.
The resulting optimal point-charge parameters are $q_{\rm Au} = 0.223(3)$, $q_{\rm Si} = 0.578(3)$, and $q_{\rm Tb} = 0.00(5)$.
The calculated inelastic spectra with the optimal charge parameters are shown by the solid lines in Fig.~\ref{figure9}(a).
The distribution of CEF levels obtained for 5000 different atomic configurations is also shown in Fig.~\ref{figure9}(b).
It can be seen that the inelastic spectra in the paramagnetic temperature range up to 50~K are reasonably reproduced by the point-charge model with the statistically distributed Au and Si atoms.
By comparing Figs.~\ref{figure9}(a) and \ref{figure9}(b), one finds that the broad peak corresponds to the transition form the quasi-degenerated ground- and first-excited states to the quasi-degenerated second- and third-excited states.
It should be emphasized that the characteristic broad peak shape of the inelastic excitation at the low temperature $T = 15$~K is well reproduced in the calculation; this indicates that the broadness is attributable to the distribution of the second excited states due to the spatial fluctuation of the local CEF.
On the other hand, it is found that such local CEF fluctuation does not affect ground state pseudo doublet, and hence, the elastic peak is still resolution limited even with the chemical disorder.

At high temperatures, we note that slight discrepancy between the calculation and observation is found in the quasielastic region around $\hbar \omega = 0$.
This may be due to the simplification used in the fitting; the intrinsic width was assumed to be negligible in the entire temperature range in order to reduce the number of adjustable parameters.
Certainly, at high temperatures the lifetime of the CEF levels shortens, and hence quasielastic tail should naturally appear in reality.

Using the optimal point-charge parameters, we estimate the principle axes of the magnetic moment distribution by diagonalizing the following expectation value matrix:
\begin{equation}
  \begin{pmatrix}
    \langle 0|J_x J_x|0 \rangle_i & \langle 0|J_x J_y|0 \rangle_i & \langle 0|J_x J_z|0 \rangle_i\\
    \langle 0|J_y J_x|0 \rangle_i & \langle 0|J_y J_y|0 \rangle_i & \langle 0|J_y J_z|0 \rangle_i\\
    \langle 0|J_z J_x|0 \rangle_i & \langle 0|J_z J_y|0 \rangle_i & \langle 0|J_z J_z|0 \rangle_i.\\
  \end{pmatrix}
\end{equation}
The eigenvector with the largest eigenvalue (i.e., largest moment direction) is obtained for each atomic configuration, and then averaged moment direction was obtained as $\langle\vec{m}\rangle \parallel (0, 0.76, -0.65)$ for the $d = 1$ site.
The direction may be compared to the magnetic moment direction $(0, 6.67, -2.63)~\mu_{\rm B}$ (for the $d = 1$ site) determined in the magnetic structure analysis.
Having the crudeness of the point-charge model in mind, we think they are in reasonable agreement showing only $20^{\circ}$ difference.

The average value of the quantization-axis component of the moment $(1/n_{\rm conf})\sum_{i} \langle 0|J_z J_z|0 \rangle_i$ is estimated as 25 (after resetting the quantization axis to the above averaged moment direction), which corresponds to $7.5~\mu_{\rm B}$ for the ordered moment.
This reduced moment size is due to the spatial fluctuation of the local CEF; some Tb$^{3+}$ has much smaller quantization axis component.
It may be noteworthy that this moment reduction is in good agreement with the ordered moment size estimated in the magnetic structure analysis, further supporting that the chemical disorder is essential in the Au$_{70}$Si$_{17}$Tb$_{13}$ approximant.

We also estimate the averaged coefficients $\bar{q}_{kq} = 1/n_{\rm conf}\sum_i q^i_{kq}$ for the spherical harmonic expansion of the CEF potential defined as Eq.~\ref{CEFexpansion}.
Dominant terms of the estimated atom-configuration-averaged $\bar{q}_{kq}$ are $\bar{q}_{2-2} = \bar{q}_{22} = -0.005$ and $\bar{q}_{20} = -0.007$.
($|\bar{q}_{kq}| < 0.002$ for other $k$ and $q$.)
It can be clearly seen that the dominant term in the spherical harmonic expansion is $\bar{q}_{20}$, which corresponds to the $B^0_2$ term in the Stevens' operator notation (see Appendix B for the relation between $q_{kq}$ and $B^q_k$).
It is well known that the point-charge model is not realistic at all, and in reality we need to take account of several other effects, such as hybridization and/or covalency.
Hence, the estimated point-charge parameters have no physical meanings.
Nevertheless, the obtained averaged coefficients $\bar{q}_{kq}$ have quantitative physical meanings, and clearly indicate that the dominant term in the CEF Hamiltonian is the second order uniaxial term ($\bar{q}_{20}$ or $B^0_2$).
This is in agreement with the earlier two CEF studies.

There is, however, stringent discrepancy in the present and the earlier studies of CEF in the quasicrystal approximant; the quantization (easy-) axis was inferred to be along the pseudo $5f$-axis in the earlier studies, whereas it is nearly perpendicular to the pseudo $5f$-axis in the present Au$_{70}$Si$_{17}$Tb$_{13}$.
In this study, both the magnetic-structure and CEF analyses consistently conclude the easy-axis direction to be nearly perpendicular to the pseudo $5f$-axis.
Therefore, we believe that the pseudo $5f$ symmetry is not the main symmetry that dominates the CEF Hamiltonian, but it is a rather weak uniaxial anisotropy in the mirror plane perpendicular to the pseudo $5f$-axis that gives rise to the dominant $B^0_2$ term.

Further note may be given by comparing the peak shape observed in the present Au$_{70}$Si$_{17}$Tb$_{13}$ approximant to the one observed in Cd$_6$Tb~\cite{DasP2017}.
We found very broad inelastic peak in the present study, and attributed it to the chemical disorder inherent to this ternary approximant.
On the other hand, the peak width is much narrower in the binary Cd$_6$Tb approximant, where the chemical disorder is less significant.
This contrasting peak widths for the CEF peaks in the binary and ternary approximant further support the decisive role of the chemical disorder for the CEF Hamiltonian in the approximant crystals.

\section{Conclusions}
Neutron elastic and inelastic scattering experiments have been performed to elucidate the microscopic magnetic properties of the quasicrystal approximant Au$_{70}$Si$_{17}$Tb$_{13}$.
Using single crystal neutron diffraction, the magnetic structure was found to be of non-collinear and non-coplanar spin order, being quite similar to the whirling spin order found in the antiferromagnetic counterpart Au$_{72}$Al$_{14}$Tb$_{14}$.
The neutron inelastic scattering clearly shows that there is only one broad magnetic-excitation peak at low temperatures.
The CEF analysis using the point-charge model taking account of the chemical disorder indicates that the dominant CEF parameter in the single-site Hamiltonian is $B^{0}_2$, representing dominant uniaxial anisotropy for the Tb$^{3+}$ in this compound.
Combining the result of magnetic structure and CEF analyses, it is concluded that the CEF gives rise to the predominantly uniaxial anisotropy axis in the local mirror plane, being nearly perpendicular to the pseudo 5f-axis, in the Au$_{70}$Si$_{17}$Tb$_{13}$ quasicrystal approximant.

\begin{acknowledgments}
  Transmission Laue diffraction characterization of the single crystal and preliminary magnetic measurements were performed using facilities of the Institute for Solid State Physics, University of Tokyo.
  The authors thank T. Haku and T. Masuda for their helps in the Laue experiment, and T. Yamauchi for the magnetic measurements.
  The authors also thank T. Sugimoto, K. Morita, and T. Tohyama for valuable discussions, and the late professor A. P. Tsai for his unparalleled supports during this work.
  Experiment at ORNL was supported by the US-Japan collaborative program on neutron scattering.
  The research at ORNL high flux isotope reactor was sponsored by the scientific user facilities division, office of basic energy sciences, US department of energy.
  The research using the HRC spectrometer was performed under the J-PARC user program with the proposal numbers 2016A0269 and 2017A0233.
  This work is partly supported by grants-in-aids for scientific research (JP15H05883, JP16H04018, JP17K18744, JP19H01834, JP19K21839, JP19H05824) from MEXT of Japan, and by the research program ``dynamic alliance for open innovation bridging human, environment, and materials.''
\end{acknowledgments}

\appendix
\section{Point charge calculation of the CEF Hamiltonian with chemical disorder}

For the present CEF calculation, the atoms within a radius of $R < R_{\rm max}$ from the center Tb$^{3+}$ site are included.
To generate list of atom positions in this range, crystallographic parameters determined in the present structure analysis (given in Table~\ref{table1}) were used.
For the point charge parameters, we assumed the same charge for the same element, regardless of the difference in the crystallographic site symmetry.
To take account of the chemical disorder resulting from the fractionally occupied sites, we assume statistical distribution of atoms at the fractional sites in a certain range $R < R_{\rm in} (< R_{\rm max})$ around the center Tb$^{3+}$ ion.
Specifically, $n_{\rm conf}$ configurations of atoms were generated in which fractional sites are occupied by either Au or Si (or vacancy for the Au8/Si8 site) with the appearance probability given by the occupancy parameter; e.g. a Au6/Si6 site in this $R < R_{\rm in}$ range is assumed to be occupied by Au with 63\% probability, or by Si with 37~\% probability.
For the atoms farer than $R \geq R_{\rm in}$, point charges at the fractionally occupied sites are assumed to be occupancy-weighted averages of individual elemental charges, e.g., $q = 0.63q_{\rm Au} + 0.37q_{\rm Si}$ for the Au6/Si6 sites in this range.

The electrostatic potential from the surrounding point charges ($q^i_j$ with $j = 1 ... p$) situated at the positions $(R_j, \theta_j, \phi_j)$ in the spherical coordinate for the $i$-th atomic configuration is then calculated as usual~\cite{JuddRD63,MulakJ00},
\begin{equation}\label{CEFpotential}
  v^i_{\rm CEF}(r, \theta, \phi) = \sum_{k = 0, 2, 4, 6} \sum_{q = -k}^{k} r^k q^i_{kq} c^{(k)}_q(\theta, \phi),
\end{equation}
where,
\begin{equation}\label{CEFexpansion}
  q^i_{kq} = \sqrt{\frac{4\pi}{2k+1}}\sum_{i = j}^{p} \frac{q^i_j c^{(k)*}_q(\theta_j, \phi_j)}{R_j^{k+1}},
\end{equation}
and $c^{(k)}_q = \sqrt{4\pi/(2k + 1)}Y^{(k)}_q(\theta, \phi)$ with $Y^{(k)}_q$ being the spherical harmonic function of the $k$-th rank.
For the CEF potential, the matrix elements $(-e) \langle J J_z|v^i_{\rm CEF}| J J_z' \rangle$ were calculated, where $e$ stands for the electron charge.
As the expectation values of the $r^n$ operator for the Tb$^{3+}$ $4f$ wavefunction, we use $\langle r^2 \rangle = 0.2302$~\AA$^{2}$, $\langle r^4 \rangle = 0.1295$~\AA$^{4}$, and $\langle r^6 \rangle = 0.1505$~\AA$^{6}$, obtained using the Dirac-Fock calculation~\cite{FreemanAJ79}.
The matrix is then diagonalized to obtain the eigenfunctions for the $i$-th configuration $|m\rangle_i$, where $m$ indexes the CEF levels.
The eigenfunctions were used in the calculation of the neutron inelastic scattering cross-sections described in the main text.

\section{Some notes on the conventions in CEF calculations}

The matrix elements of Eq.~\ref{CEFpotential} may be rewritten in terms of the spherical tensor operator (or Racah operator) $\hat{O}^{(k)}_q$ so that they coincide for the ground $J$-multiplet~\cite{SmithD66,LindgardPA74}:
\begin{equation}
  (-e)  q_{kq} \langle r^k \rangle \langle J J_z | c^{(k)}_q | J J_z' \rangle  = B^{(k)}_q \langle J J_z | \hat{O}^{(k)}_q | J J_z' \rangle.
\end{equation}
The spherical tensor operator may be obtained from its maximum state:
\begin{equation}
  \hat{O}^{(k)}_k = \frac{(-1)^k}{2^k k!}[(2 k)!]^{1/2}(\hat{J}_+)^k,
\end{equation}
and the commutation relation:
\begin{equation}
  [\hat{J}_\pm, \hat{O}^{(k)}_q] = \sqrt{k(k+1)-q(q\pm1)} \hat{O}^{(k)}_{q\pm1}.
\end{equation}
Using the reduced matrix elements for $c^{(k)}$ and $\hat{O}^{(k)}$, $B^{(k)}_q$ and $q_{kq}$ are related as:
\begin{equation}
  B^{(k)}_q = (-e) q_{kq} \langle r^k \rangle \frac{\langle J || c^{(k)} || J \rangle}{\langle J || \hat{O}^{(k)} || J \rangle}.
\end{equation}
The reduced matrix element for $\hat{O}^{(k)}$ is given as:
\begin{equation}\label{eq:Bkq}
  \langle J || \hat{O}^{(k)} || J \rangle = \frac{1}{2^k}\left [ \frac{(2 J + k + 1)!}{(2J - k)!} \right ]^{1/2}.
\end{equation}
The reduced matrix elements of $c^{(k)}$ can be calculated as prescribed in the standard text~\cite{JuddRD63,WybourneBG65}.

The Stevens' operator equivalents $\hat{O}_k^q$ used frequently in the crystal field analysis are based on the tesseral harmonics, instead of the spherical harmonics, and in addition, the some factors are dropped~\cite{HutchingsMT64,BirgeneauRJ67}.
The relation between the Racah operator equivalents $\hat{O}^{(k)}_q$ given above and the Stevens' operators $\hat{O}^q_k$ are given in Ref.~\cite{BuckmasterHA62}.


\begin{thebibliography}{46}%
\makeatletter
\providecommand \@ifxundefined [1]{%
 \@ifx{#1\undefined}
}%
\providecommand \@ifnum [1]{%
 \ifnum #1\expandafter \@firstoftwo
 \else \expandafter \@secondoftwo
 \fi
}%
\providecommand \@ifx [1]{%
 \ifx #1\expandafter \@firstoftwo
 \else \expandafter \@secondoftwo
 \fi
}%
\providecommand \natexlab [1]{#1}%
\providecommand \enquote  [1]{``#1''}%
\providecommand \bibnamefont  [1]{#1}%
\providecommand \bibfnamefont [1]{#1}%
\providecommand \citenamefont [1]{#1}%
\providecommand \href@noop [0]{\@secondoftwo}%
\providecommand \href [0]{\begingroup \@sanitize@url \@href}%
\providecommand \@href[1]{\@@startlink{#1}\@@href}%
\providecommand \@@href[1]{\endgroup#1\@@endlink}%
\providecommand \@sanitize@url [0]{\catcode `\\12\catcode `\$12\catcode
  `\&12\catcode `\#12\catcode `\^12\catcode `\_12\catcode `\%12\relax}%
\providecommand \@@startlink[1]{}%
\providecommand \@@endlink[0]{}%
\providecommand \url  [0]{\begingroup\@sanitize@url \@url }%
\providecommand \@url [1]{\endgroup\@href {#1}{\urlprefix }}%
\providecommand \urlprefix  [0]{URL }%
\providecommand \Eprint [0]{\href }%
\providecommand \doibase [0]{https://doi.org/}%
\providecommand \selectlanguage [0]{\@gobble}%
\providecommand \bibinfo  [0]{\@secondoftwo}%
\providecommand \bibfield  [0]{\@secondoftwo}%
\providecommand \translation [1]{[#1]}%
\providecommand \BibitemOpen [0]{}%
\providecommand \bibitemStop [0]{}%
\providecommand \bibitemNoStop [0]{.\EOS\space}%
\providecommand \EOS [0]{\spacefactor3000\relax}%
\providecommand \BibitemShut  [1]{\csname bibitem#1\endcsname}%
\let\auto@bib@innerbib\@empty
\bibitem [{\citenamefont {Shechtman}\ \emph {et~al.}(1984)\citenamefont
  {Shechtman}, \citenamefont {Blech}, \citenamefont {Gratias},\ and\
  \citenamefont {Cahn}}]{ShechtmanD84}%
  \BibitemOpen
  \bibfield  {author} {\bibinfo {author} {\bibfnamefont {D.}~\bibnamefont
  {Shechtman}}, \bibinfo {author} {\bibfnamefont {I.}~\bibnamefont {Blech}},
  \bibinfo {author} {\bibfnamefont {D.}~\bibnamefont {Gratias}},\ and\ \bibinfo
  {author} {\bibfnamefont {J.~W.}\ \bibnamefont {Cahn}},\ }\href
  {https://doi.org/10.1103/PhysRevLett.53.1951} {\bibfield  {journal} {\bibinfo
   {journal} {Phys. Rev. Lett.}\ }\textbf {\bibinfo {volume} {53}},\ \bibinfo
  {pages} {1951} (\bibinfo {year} {1984})}\BibitemShut {NoStop}%
\bibitem [{\citenamefont {Takakura}\ \emph {et~al.}(2007)\citenamefont
  {Takakura}, \citenamefont {Gomez}, \citenamefont {Yamamoto}, \citenamefont
  {\mbox{de} Boissieu},\ and\ \citenamefont {Tsai}}]{TakakuraH07}%
  \BibitemOpen
  \bibfield  {author} {\bibinfo {author} {\bibfnamefont {H.}~\bibnamefont
  {Takakura}}, \bibinfo {author} {\bibfnamefont {C.~P.}\ \bibnamefont {Gomez}},
  \bibinfo {author} {\bibfnamefont {A.}~\bibnamefont {Yamamoto}}, \bibinfo
  {author} {\bibfnamefont {M.}~\bibnamefont {\mbox{de} Boissieu}},\ and\
  \bibinfo {author} {\bibfnamefont {A.~P.}\ \bibnamefont {Tsai}},\ }\href@noop
  {} {\bibfield  {journal} {\bibinfo  {journal} {Nature Mater.}\ }\textbf
  {\bibinfo {volume} {6}},\ \bibinfo {pages} {58} (\bibinfo {year}
  {2007})}\BibitemShut {NoStop}%
\bibitem [{\citenamefont {Goldman}\ and\ \citenamefont
  {Kelton}(1993)}]{GoldmanAI93}%
  \BibitemOpen
  \bibfield  {author} {\bibinfo {author} {\bibfnamefont {A.~I.}\ \bibnamefont
  {Goldman}}\ and\ \bibinfo {author} {\bibfnamefont {R.~F.}\ \bibnamefont
  {Kelton}},\ }\href@noop {} {\bibfield  {journal} {\bibinfo  {journal} {Rev.
  Mod. Phys.}\ }\textbf {\bibinfo {volume} {65}},\ \bibinfo {pages} {213}
  (\bibinfo {year} {1993})}\BibitemShut {NoStop}%
\bibitem [{\citenamefont {Gomez}\ and\ \citenamefont
  {Lidin}(2003)}]{Gomez2003}%
  \BibitemOpen
  \bibfield  {author} {\bibinfo {author} {\bibfnamefont {C.~P.}\ \bibnamefont
  {Gomez}}\ and\ \bibinfo {author} {\bibfnamefont {S.}~\bibnamefont {Lidin}},\
  }\href@noop {} {\bibfield  {journal} {\bibinfo  {journal} {Phys. Rev. B}\
  }\textbf {\bibinfo {volume} {68}},\ \bibinfo {pages} {024203} (\bibinfo
  {year} {2003})}\BibitemShut {NoStop}%
\bibitem [{\citenamefont {Piao}\ \emph {et~al.}(2006)\citenamefont {Piao},
  \citenamefont {Gomez},\ and\ \citenamefont {Lidin}}]{PiaoSY06}%
  \BibitemOpen
  \bibfield  {author} {\bibinfo {author} {\bibfnamefont {S.~Y.}\ \bibnamefont
  {Piao}}, \bibinfo {author} {\bibfnamefont {C.~P.}\ \bibnamefont {Gomez}},\
  and\ \bibinfo {author} {\bibfnamefont {S.}~\bibnamefont {Lidin}},\
  }\href@noop {} {\bibfield  {journal} {\bibinfo  {journal} {Z. Naturforsch}\
  }\textbf {\bibinfo {volume} {B60}},\ \bibinfo {pages} {644} (\bibinfo {year}
  {2006})}\BibitemShut {NoStop}%
\bibitem [{\citenamefont {Bruzzone}\ \emph {et~al.}(1971)\citenamefont
  {Bruzzone}, \citenamefont {Fornasini},\ and\ \citenamefont
  {Merlo}}]{BruzzoneG71}%
  \BibitemOpen
  \bibfield  {author} {\bibinfo {author} {\bibfnamefont {G.}~\bibnamefont
  {Bruzzone}}, \bibinfo {author} {\bibfnamefont {M.}~\bibnamefont
  {Fornasini}},\ and\ \bibinfo {author} {\bibfnamefont {F.}~\bibnamefont
  {Merlo}},\ }\href
  {https://doi.org/https://doi.org/10.1016/0022-5088(71)90153-6} {\bibfield
  {journal} {\bibinfo  {journal} {J. Less Common Metals}\ }\textbf {\bibinfo
  {volume} {25}},\ \bibinfo {pages} {295 } (\bibinfo {year}
  {1971})}\BibitemShut {NoStop}%
\bibitem [{\citenamefont {Tsai}\ \emph {et~al.}(2000)\citenamefont {Tsai},
  \citenamefont {Guo}, \citenamefont {Abe}, \citenamefont {Takakura},\ and\
  \citenamefont {Sato}}]{TsaiAP00}%
  \BibitemOpen
  \bibfield  {author} {\bibinfo {author} {\bibfnamefont {A.~P.}\ \bibnamefont
  {Tsai}}, \bibinfo {author} {\bibfnamefont {J.~Q.}\ \bibnamefont {Guo}},
  \bibinfo {author} {\bibfnamefont {E.}~\bibnamefont {Abe}}, \bibinfo {author}
  {\bibfnamefont {H.}~\bibnamefont {Takakura}},\ and\ \bibinfo {author}
  {\bibfnamefont {T.~J.}\ \bibnamefont {Sato}},\ }\href@noop {} {\bibfield
  {journal} {\bibinfo  {journal} {Nature}\ }\textbf {\bibinfo {volume} {408}},\
  \bibinfo {pages} {537} (\bibinfo {year} {2000})}\BibitemShut {NoStop}%
\bibitem [{\citenamefont {Goldman}\ \emph {et~al.}(2013)\citenamefont
  {Goldman}, \citenamefont {Kong}, \citenamefont {Kreyssig}, \citenamefont
  {Jesche}, \citenamefont {Ramazanoglu}, \citenamefont {Dennis}, \citenamefont
  {Bud'ko},\ and\ \citenamefont {Canfield}}]{GoldmanAI13}%
  \BibitemOpen
  \bibfield  {author} {\bibinfo {author} {\bibfnamefont {A.~I.}\ \bibnamefont
  {Goldman}}, \bibinfo {author} {\bibfnamefont {T.}~\bibnamefont {Kong}},
  \bibinfo {author} {\bibfnamefont {A.}~\bibnamefont {Kreyssig}}, \bibinfo
  {author} {\bibfnamefont {A.}~\bibnamefont {Jesche}}, \bibinfo {author}
  {\bibfnamefont {M.}~\bibnamefont {Ramazanoglu}}, \bibinfo {author}
  {\bibfnamefont {K.~W.}\ \bibnamefont {Dennis}}, \bibinfo {author}
  {\bibfnamefont {S.~L.}\ \bibnamefont {Bud'ko}},\ and\ \bibinfo {author}
  {\bibfnamefont {P.~C.}\ \bibnamefont {Canfield}},\ }\href@noop {} {\bibfield
  {journal} {\bibinfo  {journal} {Nature Mater.}\ }\textbf {\bibinfo {volume}
  {12}},\ \bibinfo {pages} {714} (\bibinfo {year} {2013})}\BibitemShut
  {NoStop}%
\bibitem [{\citenamefont {Guo}\ \emph {et~al.}(2000)\citenamefont {Guo},
  \citenamefont {Abe},\ and\ \citenamefont {Tsai}}]{GuoJQ00}%
  \BibitemOpen
  \bibfield  {author} {\bibinfo {author} {\bibfnamefont {J.~Q.}\ \bibnamefont
  {Guo}}, \bibinfo {author} {\bibfnamefont {E.}~\bibnamefont {Abe}},\ and\
  \bibinfo {author} {\bibfnamefont {A.~P.}\ \bibnamefont {Tsai}},\ }\href@noop
  {} {\bibfield  {journal} {\bibinfo  {journal} {Jpn. J. Appl. Phys.}\ }\textbf
  {\bibinfo {volume} {39}},\ \bibinfo {pages} {L770} (\bibinfo {year}
  {2000})}\BibitemShut {NoStop}%
\bibitem [{\citenamefont {Guo}\ \emph {et~al.}(2001)\citenamefont {Guo},
  \citenamefont {Abe},\ and\ \citenamefont {Tsai}}]{GuoJQ01}%
  \BibitemOpen
  \bibfield  {author} {\bibinfo {author} {\bibfnamefont {J.~Q.}\ \bibnamefont
  {Guo}}, \bibinfo {author} {\bibfnamefont {E.}~\bibnamefont {Abe}},\ and\
  \bibinfo {author} {\bibfnamefont {A.~P.}\ \bibnamefont {Tsai}},\ }\href@noop
  {} {\bibfield  {journal} {\bibinfo  {journal} {Philos. Mag. Lett.}\ }\textbf
  {\bibinfo {volume} {81}},\ \bibinfo {pages} {17} (\bibinfo {year}
  {2001})}\BibitemShut {NoStop}%
\bibitem [{\citenamefont {Guo}\ \emph {et~al.}(2002)\citenamefont {Guo},
  \citenamefont {Abe},\ and\ \citenamefont {Tsai}}]{GuoJQ02}%
  \BibitemOpen
  \bibfield  {author} {\bibinfo {author} {\bibfnamefont {J.~Q.}\ \bibnamefont
  {Guo}}, \bibinfo {author} {\bibfnamefont {E.}~\bibnamefont {Abe}},\ and\
  \bibinfo {author} {\bibfnamefont {A.~P.}\ \bibnamefont {Tsai}},\ }\href@noop
  {} {\bibfield  {journal} {\bibinfo  {journal} {Philos. Mag. Lett.}\ }\textbf
  {\bibinfo {volume} {82}},\ \bibinfo {pages} {27} (\bibinfo {year}
  {2002})}\BibitemShut {NoStop}%
\bibitem [{\citenamefont {Tsai}\ \emph {et~al.}(1994)\citenamefont {Tsai},
  \citenamefont {Niikura}, \citenamefont {Inoue}, \citenamefont {Masumoto},
  \citenamefont {Nishita}, \citenamefont {Tsuda},\ and\ \citenamefont
  {Tanaka}}]{TsaiAP94}%
  \BibitemOpen
  \bibfield  {author} {\bibinfo {author} {\bibfnamefont {A.~P.}\ \bibnamefont
  {Tsai}}, \bibinfo {author} {\bibfnamefont {A.}~\bibnamefont {Niikura}},
  \bibinfo {author} {\bibfnamefont {A.}~\bibnamefont {Inoue}}, \bibinfo
  {author} {\bibfnamefont {T.}~\bibnamefont {Masumoto}}, \bibinfo {author}
  {\bibfnamefont {Y.}~\bibnamefont {Nishita}}, \bibinfo {author} {\bibfnamefont
  {K.}~\bibnamefont {Tsuda}},\ and\ \bibinfo {author} {\bibfnamefont
  {M.}~\bibnamefont {Tanaka}},\ }\href@noop {} {\bibfield  {journal} {\bibinfo
  {journal} {Philos. Mag. Lett.}\ }\textbf {\bibinfo {volume} {70}},\ \bibinfo
  {pages} {169} (\bibinfo {year} {1994})}\BibitemShut {NoStop}%
\bibitem [{\citenamefont {Morita}\ and\ \citenamefont
  {Tsai}(2008)}]{MoritaY08}%
  \BibitemOpen
  \bibfield  {author} {\bibinfo {author} {\bibfnamefont {Y.}~\bibnamefont
  {Morita}}\ and\ \bibinfo {author} {\bibfnamefont {A.~P.}\ \bibnamefont
  {Tsai}},\ }\href@noop {} {\bibfield  {journal} {\bibinfo  {journal} {Jpn. J.
  Appl. Phys.}\ }\textbf {\bibinfo {volume} {47}},\ \bibinfo {pages} {7975}
  (\bibinfo {year} {2008})}\BibitemShut {NoStop}%
\bibitem [{\citenamefont {Kenzari}\ \emph {et~al.}(2008)\citenamefont
  {Kenzari}, \citenamefont {Demange}, \citenamefont {Boulet}, \citenamefont
  {de~Weerd}, \citenamefont {Ledieu}, \citenamefont {Dubois},\ and\
  \citenamefont {Fourn{\'{e}}e}}]{KenzariS08}%
  \BibitemOpen
  \bibfield  {author} {\bibinfo {author} {\bibfnamefont {S.}~\bibnamefont
  {Kenzari}}, \bibinfo {author} {\bibfnamefont {V.}~\bibnamefont {Demange}},
  \bibinfo {author} {\bibfnamefont {P.}~\bibnamefont {Boulet}}, \bibinfo
  {author} {\bibfnamefont {M.~C.}\ \bibnamefont {de~Weerd}}, \bibinfo {author}
  {\bibfnamefont {J.}~\bibnamefont {Ledieu}}, \bibinfo {author} {\bibfnamefont
  {J.~M.}\ \bibnamefont {Dubois}},\ and\ \bibinfo {author} {\bibfnamefont
  {V.}~\bibnamefont {Fourn{\'{e}}e}},\ }\href
  {https://doi.org/10.1088/0953-8984/20/9/095218} {\bibfield  {journal}
  {\bibinfo  {journal} {J. Phys.: Condens. Matter}\ }\textbf {\bibinfo {volume}
  {20}},\ \bibinfo {pages} {095218} (\bibinfo {year} {2008})}\BibitemShut
  {NoStop}%
\bibitem [{\citenamefont {Gebresenbut}\ \emph {et~al.}(2013)\citenamefont
  {Gebresenbut}, \citenamefont {Tamura}, \citenamefont {Ekl\"{o}f},\ and\
  \citenamefont {G{\'{o}}mez}}]{GebresenbutGH13}%
  \BibitemOpen
  \bibfield  {author} {\bibinfo {author} {\bibfnamefont {G.~H.}\ \bibnamefont
  {Gebresenbut}}, \bibinfo {author} {\bibfnamefont {R.}~\bibnamefont {Tamura}},
  \bibinfo {author} {\bibfnamefont {D.}~\bibnamefont {Ekl\"{o}f}},\ and\
  \bibinfo {author} {\bibfnamefont {C.~P.}\ \bibnamefont {G{\'{o}}mez}},\
  }\href {https://doi.org/10.1088/0953-8984/25/13/135402} {\bibfield  {journal}
  {\bibinfo  {journal} {J. Phys.: Condens. Matter}\ }\textbf {\bibinfo {volume}
  {25}},\ \bibinfo {pages} {135402} (\bibinfo {year} {2013})}\BibitemShut
  {NoStop}%
\bibitem [{\citenamefont {Tamura}\ \emph {et~al.}(2002)\citenamefont {Tamura},
  \citenamefont {Murao}, \citenamefont {Takeuchi}, \citenamefont {Ichihara},
  \citenamefont {Isobe},\ and\ \citenamefont {Ueda}}]{Tamura2002}%
  \BibitemOpen
  \bibfield  {author} {\bibinfo {author} {\bibfnamefont {R.}~\bibnamefont
  {Tamura}}, \bibinfo {author} {\bibfnamefont {Y.}~\bibnamefont {Murao}},
  \bibinfo {author} {\bibfnamefont {S.}~\bibnamefont {Takeuchi}}, \bibinfo
  {author} {\bibfnamefont {M.}~\bibnamefont {Ichihara}}, \bibinfo {author}
  {\bibfnamefont {M.}~\bibnamefont {Isobe}},\ and\ \bibinfo {author}
  {\bibfnamefont {Y.}~\bibnamefont {Ueda}},\ }\href@noop {} {\bibfield
  {journal} {\bibinfo  {journal} {Jpn. J. Appl. Phys.}\ }\textbf {\bibinfo
  {volume} {41}},\ \bibinfo {pages} {L524} (\bibinfo {year}
  {2002})}\BibitemShut {NoStop}%
\bibitem [{\citenamefont {Tamura}\ \emph {et~al.}(2005)\citenamefont {Tamura},
  \citenamefont {Edagawa}, \citenamefont {Shibata}, \citenamefont {Nishimoto},
  \citenamefont {Takeuchi}, \citenamefont {Saitoh}, \citenamefont {Isobe},\
  and\ \citenamefont {Ueda}}]{TamuraR05}%
  \BibitemOpen
  \bibfield  {author} {\bibinfo {author} {\bibfnamefont {R.}~\bibnamefont
  {Tamura}}, \bibinfo {author} {\bibfnamefont {K.}~\bibnamefont {Edagawa}},
  \bibinfo {author} {\bibfnamefont {K.}~\bibnamefont {Shibata}}, \bibinfo
  {author} {\bibfnamefont {K.}~\bibnamefont {Nishimoto}}, \bibinfo {author}
  {\bibfnamefont {S.}~\bibnamefont {Takeuchi}}, \bibinfo {author}
  {\bibfnamefont {K.}~\bibnamefont {Saitoh}}, \bibinfo {author} {\bibfnamefont
  {M.}~\bibnamefont {Isobe}},\ and\ \bibinfo {author} {\bibfnamefont
  {Y.}~\bibnamefont {Ueda}},\ }\href
  {https://doi.org/10.1103/PhysRevB.72.174211} {\bibfield  {journal} {\bibinfo
  {journal} {Phys. Rev. B}\ }\textbf {\bibinfo {volume} {72}},\ \bibinfo
  {pages} {174211} (\bibinfo {year} {2005})}\BibitemShut {NoStop}%
\bibitem [{\citenamefont {Nishimoto}\ \emph {et~al.}(2013)\citenamefont
  {Nishimoto}, \citenamefont {Sato},\ and\ \citenamefont
  {Tamura}}]{NishimotoK13}%
  \BibitemOpen
  \bibfield  {author} {\bibinfo {author} {\bibfnamefont {K.}~\bibnamefont
  {Nishimoto}}, \bibinfo {author} {\bibfnamefont {T.}~\bibnamefont {Sato}},\
  and\ \bibinfo {author} {\bibfnamefont {R.}~\bibnamefont {Tamura}},\ }\href
  {https://doi.org/10.1088/0953-8984/25/23/235403} {\bibfield  {journal}
  {\bibinfo  {journal} {J. Phys.: Condens. Matter}\ }\textbf {\bibinfo {volume}
  {25}},\ \bibinfo {pages} {235403} (\bibinfo {year} {2013})}\BibitemShut
  {NoStop}%
\bibitem [{\citenamefont {Tamura}\ \emph {et~al.}(2010)\citenamefont {Tamura},
  \citenamefont {Muro}, \citenamefont {Hiroto}, \citenamefont {Nishimoto}, ,\
  and\ \citenamefont {Takabatake}}]{Tamura2010}%
  \BibitemOpen
  \bibfield  {author} {\bibinfo {author} {\bibfnamefont {R.}~\bibnamefont
  {Tamura}}, \bibinfo {author} {\bibfnamefont {Y.}~\bibnamefont {Muro}},
  \bibinfo {author} {\bibfnamefont {T.}~\bibnamefont {Hiroto}}, \bibinfo
  {author} {\bibfnamefont {K.}~\bibnamefont {Nishimoto}}, ,\ and\ \bibinfo
  {author} {\bibfnamefont {T.}~\bibnamefont {Takabatake}},\ }\href@noop {}
  {\bibfield  {journal} {\bibinfo  {journal} {Phys. Rev. B}\ }\textbf {\bibinfo
  {volume} {82}},\ \bibinfo {pages} {220201(R)} (\bibinfo {year}
  {2010})}\BibitemShut {NoStop}%
\bibitem [{\citenamefont {Mori}\ \emph {et~al.}(2012)\citenamefont {Mori},
  \citenamefont {Ota}, \citenamefont {Yoshiuchi}, \citenamefont {Iwakawa},
  \citenamefont {Taga}, \citenamefont {Hirose}, \citenamefont {Takeuchi},
  \citenamefont {Yamamoto}, \citenamefont {Haga}, \citenamefont {Honda},
  \citenamefont {Settai},\ and\ \citenamefont {Onuki}}]{MoriA12}%
  \BibitemOpen
  \bibfield  {author} {\bibinfo {author} {\bibfnamefont {A.}~\bibnamefont
  {Mori}}, \bibinfo {author} {\bibfnamefont {H.}~\bibnamefont {Ota}}, \bibinfo
  {author} {\bibfnamefont {S.}~\bibnamefont {Yoshiuchi}}, \bibinfo {author}
  {\bibfnamefont {K.}~\bibnamefont {Iwakawa}}, \bibinfo {author} {\bibfnamefont
  {Y.}~\bibnamefont {Taga}}, \bibinfo {author} {\bibfnamefont {Y.}~\bibnamefont
  {Hirose}}, \bibinfo {author} {\bibfnamefont {T.}~\bibnamefont {Takeuchi}},
  \bibinfo {author} {\bibfnamefont {E.}~\bibnamefont {Yamamoto}}, \bibinfo
  {author} {\bibfnamefont {Y.}~\bibnamefont {Haga}}, \bibinfo {author}
  {\bibfnamefont {F.}~\bibnamefont {Honda}}, \bibinfo {author} {\bibfnamefont
  {R.}~\bibnamefont {Settai}},\ and\ \bibinfo {author} {\bibfnamefont
  {Y.}~\bibnamefont {Onuki}},\ }\href {https://doi.org/10.1143/JPSJ.81.024720}
  {\bibfield  {journal} {\bibinfo  {journal} {J. Phys. Soc. Jpn.}\ }\textbf
  {\bibinfo {volume} {81}},\ \bibinfo {pages} {024720} (\bibinfo {year}
  {2012})}\BibitemShut {NoStop}%
\bibitem [{\citenamefont {Tamura}\ \emph {et~al.}(2012)\citenamefont {Tamura},
  \citenamefont {Muro}, \citenamefont {Hiroto}, \citenamefont {Yaguchi},
  \citenamefont {Beutier},\ and\ \citenamefont {Takabatake}}]{TamuraR12}%
  \BibitemOpen
  \bibfield  {author} {\bibinfo {author} {\bibfnamefont {R.}~\bibnamefont
  {Tamura}}, \bibinfo {author} {\bibfnamefont {Y.}~\bibnamefont {Muro}},
  \bibinfo {author} {\bibfnamefont {T.}~\bibnamefont {Hiroto}}, \bibinfo
  {author} {\bibfnamefont {H.}~\bibnamefont {Yaguchi}}, \bibinfo {author}
  {\bibfnamefont {G.}~\bibnamefont {Beutier}},\ and\ \bibinfo {author}
  {\bibfnamefont {T.}~\bibnamefont {Takabatake}},\ }\href
  {https://doi.org/10.1103/PhysRevB.85.014203} {\bibfield  {journal} {\bibinfo
  {journal} {Phys. Rev. B}\ }\textbf {\bibinfo {volume} {85}},\ \bibinfo
  {pages} {014203} (\bibinfo {year} {2012})}\BibitemShut {NoStop}%
\bibitem [{\citenamefont {Hiroto}\ \emph {et~al.}(2013)\citenamefont {Hiroto},
  \citenamefont {Gebresenbut}, \citenamefont {G{\'{o}}mez}, \citenamefont
  {Muro}, \citenamefont {Isobe}, \citenamefont {Ueda}, \citenamefont {Tokiwa},\
  and\ \citenamefont {Tamura}}]{Hiroto2013}%
  \BibitemOpen
  \bibfield  {author} {\bibinfo {author} {\bibfnamefont {T.}~\bibnamefont
  {Hiroto}}, \bibinfo {author} {\bibfnamefont {G.~H.}\ \bibnamefont
  {Gebresenbut}}, \bibinfo {author} {\bibfnamefont {C.~P.}\ \bibnamefont
  {G{\'{o}}mez}}, \bibinfo {author} {\bibfnamefont {Y.}~\bibnamefont {Muro}},
  \bibinfo {author} {\bibfnamefont {M.}~\bibnamefont {Isobe}}, \bibinfo
  {author} {\bibfnamefont {Y.}~\bibnamefont {Ueda}}, \bibinfo {author}
  {\bibfnamefont {K.}~\bibnamefont {Tokiwa}},\ and\ \bibinfo {author}
  {\bibfnamefont {R.}~\bibnamefont {Tamura}},\ }\href
  {https://doi.org/10.1088/0953-8984/25/42/426004} {\bibfield  {journal}
  {\bibinfo  {journal} {J. Phys.: Condens. Matter}\ }\textbf {\bibinfo {volume}
  {25}},\ \bibinfo {pages} {426004} (\bibinfo {year} {2013})}\BibitemShut
  {NoStop}%
\bibitem [{\citenamefont {Hiroto}\ \emph {et~al.}(2014)\citenamefont {Hiroto},
  \citenamefont {Tokiwa},\ and\ \citenamefont {Tamura}}]{HirotoT14}%
  \BibitemOpen
  \bibfield  {author} {\bibinfo {author} {\bibfnamefont {T.}~\bibnamefont
  {Hiroto}}, \bibinfo {author} {\bibfnamefont {K.}~\bibnamefont {Tokiwa}},\
  and\ \bibinfo {author} {\bibfnamefont {R.}~\bibnamefont {Tamura}},\ }\href
  {https://doi.org/10.1088/0953-8984/26/21/216004} {\bibfield  {journal}
  {\bibinfo  {journal} {J. Phys.: Condens. Matter}\ }\textbf {\bibinfo {volume}
  {26}},\ \bibinfo {pages} {216004} (\bibinfo {year} {2014})}\BibitemShut
  {NoStop}%
\bibitem [{\citenamefont {Kim}\ \emph {et~al.}(2012)\citenamefont {Kim},
  \citenamefont {Beutier}, \citenamefont {Kreyssig}, \citenamefont {Hiroto},
  \citenamefont {Yamada}, \citenamefont {Kim}, \citenamefont {de~Boissieu},
  \citenamefont {Tamura},\ and\ \citenamefont {Goldman}}]{Kim2012}%
  \BibitemOpen
  \bibfield  {author} {\bibinfo {author} {\bibfnamefont {M.~G.}\ \bibnamefont
  {Kim}}, \bibinfo {author} {\bibfnamefont {G.}~\bibnamefont {Beutier}},
  \bibinfo {author} {\bibfnamefont {A.}~\bibnamefont {Kreyssig}}, \bibinfo
  {author} {\bibfnamefont {T.}~\bibnamefont {Hiroto}}, \bibinfo {author}
  {\bibfnamefont {T.}~\bibnamefont {Yamada}}, \bibinfo {author} {\bibfnamefont
  {J.~W.}\ \bibnamefont {Kim}}, \bibinfo {author} {\bibfnamefont
  {M.}~\bibnamefont {de~Boissieu}}, \bibinfo {author} {\bibfnamefont
  {R.}~\bibnamefont {Tamura}},\ and\ \bibinfo {author} {\bibfnamefont {A.~I.}\
  \bibnamefont {Goldman}},\ }\href@noop {} {\bibfield  {journal} {\bibinfo
  {journal} {Phys. Rev. B}\ }\textbf {\bibinfo {volume} {85}},\ \bibinfo
  {pages} {134442} (\bibinfo {year} {2012})}\BibitemShut {NoStop}%
\bibitem [{\citenamefont {Kreyssig}\ \emph {et~al.}(2013)\citenamefont
  {Kreyssig}, \citenamefont {Beutier}, \citenamefont {Hiroto}, \citenamefont
  {Kim}, \citenamefont {Tucker}, \citenamefont {\mbox{M. de} Boissieu~adn
  R.~Tamura},\ and\ \citenamefont {Goldman}}]{KreyssigA13}%
  \BibitemOpen
  \bibfield  {author} {\bibinfo {author} {\bibfnamefont {A.}~\bibnamefont
  {Kreyssig}}, \bibinfo {author} {\bibfnamefont {G.}~\bibnamefont {Beutier}},
  \bibinfo {author} {\bibfnamefont {T.}~\bibnamefont {Hiroto}}, \bibinfo
  {author} {\bibfnamefont {M.~G.}\ \bibnamefont {Kim}}, \bibinfo {author}
  {\bibfnamefont {G.~S.}\ \bibnamefont {Tucker}}, \bibinfo {author}
  {\bibnamefont {\mbox{M. de} Boissieu~adn R.~Tamura}},\ and\ \bibinfo {author}
  {\bibfnamefont {A.~I.}\ \bibnamefont {Goldman}},\ }\href@noop {} {\bibfield
  {journal} {\bibinfo  {journal} {Philos. Mag. Lett.}\ }\textbf {\bibinfo
  {volume} {93}},\ \bibinfo {pages} {512} (\bibinfo {year} {2013})}\BibitemShut
  {NoStop}%
\bibitem [{\citenamefont {Gebresenbut}\ \emph {et~al.}(2014)\citenamefont
  {Gebresenbut}, \citenamefont {Andersson}, \citenamefont {Beran},
  \citenamefont {Manuel}, \citenamefont {Nordblad}, \citenamefont {Sahlberg},\
  and\ \citenamefont {Gomez}}]{Gebresenbut2014}%
  \BibitemOpen
  \bibfield  {author} {\bibinfo {author} {\bibfnamefont {G.}~\bibnamefont
  {Gebresenbut}}, \bibinfo {author} {\bibfnamefont {M.~S.}\ \bibnamefont
  {Andersson}}, \bibinfo {author} {\bibfnamefont {P.}~\bibnamefont {Beran}},
  \bibinfo {author} {\bibfnamefont {P.}~\bibnamefont {Manuel}}, \bibinfo
  {author} {\bibfnamefont {P.}~\bibnamefont {Nordblad}}, \bibinfo {author}
  {\bibfnamefont {M.}~\bibnamefont {Sahlberg}},\ and\ \bibinfo {author}
  {\bibfnamefont {C.~P.}\ \bibnamefont {Gomez}},\ }\href
  {https://doi.org/10.1088/0953-8984/26/32/322202} {\bibfield  {journal}
  {\bibinfo  {journal} {J. Phys.: Condens. Matter}\ }\textbf {\bibinfo {volume}
  {26}},\ \bibinfo {pages} {322202} (\bibinfo {year} {2014})}\BibitemShut
  {NoStop}%
\bibitem [{\citenamefont {Gebresenbut}()}]{GebresenbutGH19}%
  \BibitemOpen
  \bibfield  {author} {\bibinfo {author} {\bibfnamefont {G.~H.}\ \bibnamefont
  {Gebresenbut}},\ }\bibinfo {note} {presented at the 14th International
  Conference on Quasicrystals (Kranjska Gora, Slovenia, 26-31 May
  2019)}\BibitemShut {NoStop}%
\bibitem [{\citenamefont {Sugimoto}\ \emph {et~al.}(2016)\citenamefont
  {Sugimoto}, \citenamefont {Tohyama}, \citenamefont {Hiroto},\ and\
  \citenamefont {Tamura}}]{SugimotoT16}%
  \BibitemOpen
  \bibfield  {author} {\bibinfo {author} {\bibfnamefont {T.}~\bibnamefont
  {Sugimoto}}, \bibinfo {author} {\bibfnamefont {T.}~\bibnamefont {Tohyama}},
  \bibinfo {author} {\bibfnamefont {T.}~\bibnamefont {Hiroto}},\ and\ \bibinfo
  {author} {\bibfnamefont {R.}~\bibnamefont {Tamura}},\ }\href
  {https://doi.org/10.7566/JPSJ.85.053701} {\bibfield  {journal} {\bibinfo
  {journal} {J. Phys. Soc. Jpn.}\ }\textbf {\bibinfo {volume} {85}},\ \bibinfo
  {pages} {053701} (\bibinfo {year} {2016})}\BibitemShut {NoStop}%
\bibitem [{\citenamefont {Sato}\ \emph {et~al.}(2019)\citenamefont {Sato},
  \citenamefont {Ishikawa}, \citenamefont {Sakurai}, \citenamefont {Hattori},
  \citenamefont {Avdeev},\ and\ \citenamefont {Tamura}}]{SatoTJ19}%
  \BibitemOpen
  \bibfield  {author} {\bibinfo {author} {\bibfnamefont {T.~J.}\ \bibnamefont
  {Sato}}, \bibinfo {author} {\bibfnamefont {A.}~\bibnamefont {Ishikawa}},
  \bibinfo {author} {\bibfnamefont {A.}~\bibnamefont {Sakurai}}, \bibinfo
  {author} {\bibfnamefont {M.}~\bibnamefont {Hattori}}, \bibinfo {author}
  {\bibfnamefont {M.}~\bibnamefont {Avdeev}},\ and\ \bibinfo {author}
  {\bibfnamefont {R.}~\bibnamefont {Tamura}},\ }\href
  {https://doi.org/10.1103/PhysRevB.100.054417} {\bibfield  {journal} {\bibinfo
   {journal} {Phys. Rev. B}\ }\textbf {\bibinfo {volume} {100}},\ \bibinfo
  {pages} {054417} (\bibinfo {year} {2019})}\BibitemShut {NoStop}%
\bibitem [{\citenamefont {Jazbec}\ \emph {et~al.}(2016)\citenamefont {Jazbec},
  \citenamefont {Kashimoto}, \citenamefont {Ko\ifmmode~\check{z}\else
  \v{z}\fi{}elj}, \citenamefont {Vrtnik}, \citenamefont
  {Jagodi\ifmmode~\check{c}\else \v{c}\fi{}}, \citenamefont {Jagli\ifmmode
  \check{c}\else \v{c}\fi{}i\ifmmode~\acute{c}\else \'{c}\fi{}},\ and\
  \citenamefont {Dolin\ifmmode~\check{s}\else \v{s}\fi{}ek}}]{JazbecS16}%
  \BibitemOpen
  \bibfield  {author} {\bibinfo {author} {\bibfnamefont {S.}~\bibnamefont
  {Jazbec}}, \bibinfo {author} {\bibfnamefont {S.}~\bibnamefont {Kashimoto}},
  \bibinfo {author} {\bibfnamefont {P.}~\bibnamefont {Ko\ifmmode~\check{z}\else
  \v{z}\fi{}elj}}, \bibinfo {author} {\bibfnamefont {S.}~\bibnamefont
  {Vrtnik}}, \bibinfo {author} {\bibfnamefont {M.}~\bibnamefont
  {Jagodi\ifmmode~\check{c}\else \v{c}\fi{}}}, \bibinfo {author} {\bibfnamefont
  {Z.}~\bibnamefont {Jagli\ifmmode \check{c}\else
  \v{c}\fi{}i\ifmmode~\acute{c}\else \'{c}\fi{}}},\ and\ \bibinfo {author}
  {\bibfnamefont {J.}~\bibnamefont {Dolin\ifmmode~\check{s}\else
  \v{s}\fi{}ek}},\ }\href {https://doi.org/10.1103/PhysRevB.93.054208}
  {\bibfield  {journal} {\bibinfo  {journal} {Phys. Rev. B}\ }\textbf {\bibinfo
  {volume} {93}},\ \bibinfo {pages} {054208} (\bibinfo {year}
  {2016})}\BibitemShut {NoStop}%
\bibitem [{\citenamefont {Das}\ \emph {et~al.}(2017)\citenamefont {Das},
  \citenamefont {Lory}, \citenamefont {Flint}, \citenamefont {Kong},
  \citenamefont {Hiroto}, \citenamefont {Bud'ko}, \citenamefont {Canfield},
  \citenamefont {de~Boissieu}, \citenamefont {Kreyssig},\ and\ \citenamefont
  {Goldman}}]{DasP2017}%
  \BibitemOpen
  \bibfield  {author} {\bibinfo {author} {\bibfnamefont {P.}~\bibnamefont
  {Das}}, \bibinfo {author} {\bibfnamefont {P.-F.}\ \bibnamefont {Lory}},
  \bibinfo {author} {\bibfnamefont {R.}~\bibnamefont {Flint}}, \bibinfo
  {author} {\bibfnamefont {T.}~\bibnamefont {Kong}}, \bibinfo {author}
  {\bibfnamefont {T.}~\bibnamefont {Hiroto}}, \bibinfo {author} {\bibfnamefont
  {S.~L.}\ \bibnamefont {Bud'ko}}, \bibinfo {author} {\bibfnamefont {P.~C.}\
  \bibnamefont {Canfield}}, \bibinfo {author} {\bibfnamefont {M.}~\bibnamefont
  {de~Boissieu}}, \bibinfo {author} {\bibfnamefont {A.}~\bibnamefont
  {Kreyssig}},\ and\ \bibinfo {author} {\bibfnamefont {A.~I.}\ \bibnamefont
  {Goldman}},\ }\href {https://doi.org/10.1103/PhysRevB.95.054408} {\bibfield
  {journal} {\bibinfo  {journal} {Phys. Rev. B}\ }\textbf {\bibinfo {volume}
  {95}},\ \bibinfo {pages} {054408} (\bibinfo {year} {2017})}\BibitemShut
  {NoStop}%
\bibitem [{\citenamefont {Chakoumakos}\ \emph {et~al.}(2011)\citenamefont
  {Chakoumakos}, \citenamefont {Cao}, \citenamefont {Ye}, \citenamefont
  {Stoica}, \citenamefont {Popovici}, \citenamefont {Sundaram}, \citenamefont
  {Zhou}, \citenamefont {Hicks}, \citenamefont {Lynn},\ and\ \citenamefont
  {Riedel}}]{ChakoumakosBC11}%
  \BibitemOpen
  \bibfield  {author} {\bibinfo {author} {\bibfnamefont {B.~C.}\ \bibnamefont
  {Chakoumakos}}, \bibinfo {author} {\bibfnamefont {H.}~\bibnamefont {Cao}},
  \bibinfo {author} {\bibfnamefont {F.}~\bibnamefont {Ye}}, \bibinfo {author}
  {\bibfnamefont {A.~D.}\ \bibnamefont {Stoica}}, \bibinfo {author}
  {\bibfnamefont {M.}~\bibnamefont {Popovici}}, \bibinfo {author}
  {\bibfnamefont {M.}~\bibnamefont {Sundaram}}, \bibinfo {author}
  {\bibfnamefont {W.}~\bibnamefont {Zhou}}, \bibinfo {author} {\bibfnamefont
  {J.~S.}\ \bibnamefont {Hicks}}, \bibinfo {author} {\bibfnamefont {G.~W.}\
  \bibnamefont {Lynn}},\ and\ \bibinfo {author} {\bibfnamefont {R.~A.}\
  \bibnamefont {Riedel}},\ }\href {https://doi.org/10.1107/S0021889811012301}
  {\bibfield  {journal} {\bibinfo  {journal} {J. Appl. Crystallogr.}\ }\textbf
  {\bibinfo {volume} {44}},\ \bibinfo {pages} {655} (\bibinfo {year}
  {2011})}\BibitemShut {NoStop}%
\bibitem [{MSA()}]{MSAS2019}%
  \BibitemOpen
  \href@noop {} {}\bibinfo {note} {\noindent
  http://www2.tagen.tohoku.ac.jp/lab/sato\_tj/magnetic-representations-and-magnetic-space-groups/}\BibitemShut
  {NoStop}%
\bibitem [{\citenamefont {Itoh}\ \emph {et~al.}(2011)\citenamefont {Itoh},
  \citenamefont {Yokoo}, \citenamefont {Satoh}, \citenamefont {Yano},
  \citenamefont {Kawana}, \citenamefont {Suzuki},\ and\ \citenamefont
  {Sato}}]{ItohS11}%
  \BibitemOpen
  \bibfield  {author} {\bibinfo {author} {\bibfnamefont {S.}~\bibnamefont
  {Itoh}}, \bibinfo {author} {\bibfnamefont {T.}~\bibnamefont {Yokoo}},
  \bibinfo {author} {\bibfnamefont {S.}~\bibnamefont {Satoh}}, \bibinfo
  {author} {\bibfnamefont {S.}~\bibnamefont {Yano}}, \bibinfo {author}
  {\bibfnamefont {D.}~\bibnamefont {Kawana}}, \bibinfo {author} {\bibfnamefont
  {J.}~\bibnamefont {Suzuki}},\ and\ \bibinfo {author} {\bibfnamefont {T.~J.}\
  \bibnamefont {Sato}},\ }\href
  {https://doi.org/https://doi.org/10.1016/j.nima.2010.11.107} {\bibfield
  {journal} {\bibinfo  {journal} {Nucl. Instr. Meth.}\ }\textbf {\bibinfo
  {volume} {A631}},\ \bibinfo {pages} {90 } (\bibinfo {year}
  {2011})}\BibitemShut {NoStop}%
\bibitem [{\citenamefont {Gebresenbut}\ \emph {et~al.}(2016)\citenamefont
  {Gebresenbut}, \citenamefont {Andersson}, \citenamefont {Nordblad},
  \citenamefont {Sahlberg},\ and\ \citenamefont {Gomez}}]{GebresenbutGH16}%
  \BibitemOpen
  \bibfield  {author} {\bibinfo {author} {\bibfnamefont {G.~H.}\ \bibnamefont
  {Gebresenbut}}, \bibinfo {author} {\bibfnamefont {M.~S.}\ \bibnamefont
  {Andersson}}, \bibinfo {author} {\bibfnamefont {P.}~\bibnamefont {Nordblad}},
  \bibinfo {author} {\bibfnamefont {M.}~\bibnamefont {Sahlberg}},\ and\
  \bibinfo {author} {\bibfnamefont {C.~P.}\ \bibnamefont {Gomez}},\ }\href@noop
  {} {\bibfield  {journal} {\bibinfo  {journal} {Inorg. Chem.}\ }\textbf
  {\bibinfo {volume} {55}},\ \bibinfo {pages} {2001} (\bibinfo {year}
  {2016})}\BibitemShut {NoStop}%
\bibitem [{\citenamefont {Izyumov}\ and\ \citenamefont
  {Naish}(1979)}]{Izyumov1979}%
  \BibitemOpen
  \bibfield  {author} {\bibinfo {author} {\bibfnamefont {Y.}~\bibnamefont
  {Izyumov}}\ and\ \bibinfo {author} {\bibfnamefont {V.}~\bibnamefont
  {Naish}},\ }\href
  {https://doi.org/https://doi.org/10.1016/0304-8853(79)90086-6} {\bibfield
  {journal} {\bibinfo  {journal} {J. Magn. Magn. Mater.}\ }\textbf {\bibinfo
  {volume} {12}},\ \bibinfo {pages} {239 } (\bibinfo {year}
  {1979})}\BibitemShut {NoStop}%
\bibitem [{\citenamefont {Izyumov}\ \emph {et~al.}(1991)\citenamefont
  {Izyumov}, \citenamefont {Naish},\ and\ \citenamefont
  {Ozerov}}]{Izyumov1991}%
  \BibitemOpen
  \bibfield  {author} {\bibinfo {author} {\bibfnamefont {Y.~A.}\ \bibnamefont
  {Izyumov}}, \bibinfo {author} {\bibfnamefont {V.~E.}\ \bibnamefont {Naish}},\
  and\ \bibinfo {author} {\bibfnamefont {R.~P.}\ \bibnamefont {Ozerov}},\
  }\href@noop {} {\emph {\bibinfo {title} {Neutron Diffraction of Magnetic
  Materials}}}\ (\bibinfo  {publisher} {Springer Berlin Heidelberg},\ \bibinfo
  {address} {Berlin, Heidelberg},\ \bibinfo {year} {1991})\BibitemShut
  {NoStop}%
\bibitem [{\citenamefont {Freeman}\ and\ \citenamefont
  {Desclaux}(1979)}]{FreemanAJ79}%
  \BibitemOpen
  \bibfield  {author} {\bibinfo {author} {\bibfnamefont {A.}~\bibnamefont
  {Freeman}}\ and\ \bibinfo {author} {\bibfnamefont {J.}~\bibnamefont
  {Desclaux}},\ }\href
  {https://doi.org/https://doi.org/10.1016/0304-8853(79)90328-7} {\bibfield
  {journal} {\bibinfo  {journal} {J. Magn. Magn. Mater.}\ }\textbf {\bibinfo
  {volume} {12}},\ \bibinfo {pages} {11 } (\bibinfo {year} {1979})}\BibitemShut
  {NoStop}%
\bibitem [{\citenamefont {Hutchings}(1964)}]{HutchingsMT64}%
  \BibitemOpen
  \bibfield  {author} {\bibinfo {author} {\bibfnamefont {M.~T.}\ \bibnamefont
  {Hutchings}},\ }\href@noop {} {\bibfield  {journal} {\bibinfo  {journal}
  {Solid State Phys.}\ }\textbf {\bibinfo {volume} {16}},\ \bibinfo {pages}
  {227} (\bibinfo {year} {1964})}\BibitemShut {NoStop}%
\bibitem [{\citenamefont {Judd}(1963)}]{JuddRD63}%
  \BibitemOpen
  \bibfield  {author} {\bibinfo {author} {\bibfnamefont {B.~R.}\ \bibnamefont
  {Judd}},\ }\href@noop {} {\emph {\bibinfo {title} {Operator Techniques in
  Atomic Spectroscopy}}}\ (\bibinfo  {publisher} {McGraw-Hill},\ \bibinfo
  {address} {New York},\ \bibinfo {year} {1963})\BibitemShut {NoStop}%
\bibitem [{\citenamefont {Mulak}\ and\ \citenamefont {Gajek}(2000)}]{MulakJ00}%
  \BibitemOpen
  \bibfield  {author} {\bibinfo {author} {\bibfnamefont {J.}~\bibnamefont
  {Mulak}}\ and\ \bibinfo {author} {\bibfnamefont {Z.}~\bibnamefont {Gajek}},\
  }\href@noop {} {\emph {\bibinfo {title} {The Effective Crystal Field
  Potential}}}\ (\bibinfo  {publisher} {Elsevier},\ \bibinfo {address}
  {Amsterdam},\ \bibinfo {year} {2000})\BibitemShut {NoStop}%
\bibitem [{\citenamefont {Smith}\ and\ \citenamefont
  {Thornley}(1966)}]{SmithD66}%
  \BibitemOpen
  \bibfield  {author} {\bibinfo {author} {\bibfnamefont {D.}~\bibnamefont
  {Smith}}\ and\ \bibinfo {author} {\bibfnamefont {J.~H.~M.}\ \bibnamefont
  {Thornley}},\ }\href@noop {} {\bibfield  {journal} {\bibinfo  {journal}
  {Proc. Phys. Soc.}\ }\textbf {\bibinfo {volume} {89}},\ \bibinfo {pages}
  {779} (\bibinfo {year} {1966})}\BibitemShut {NoStop}%
\bibitem [{\citenamefont {Lindgard}\ and\ \citenamefont
  {Danielsen}(1974)}]{LindgardPA74}%
  \BibitemOpen
  \bibfield  {author} {\bibinfo {author} {\bibfnamefont {P.-A.}\ \bibnamefont
  {Lindgard}}\ and\ \bibinfo {author} {\bibfnamefont {O.}~\bibnamefont
  {Danielsen}},\ }\href@noop {} {\bibfield  {journal} {\bibinfo  {journal} {J.
  Phys. C: Solid State Phys.}\ }\textbf {\bibinfo {volume} {7}},\ \bibinfo
  {pages} {1523} (\bibinfo {year} {1974})}\BibitemShut {NoStop}%
\bibitem [{\citenamefont {Wybourne}(1965)}]{WybourneBG65}%
  \BibitemOpen
  \bibfield  {author} {\bibinfo {author} {\bibfnamefont {B.~G.}\ \bibnamefont
  {Wybourne}},\ }\href@noop {} {\emph {\bibinfo {title} {Spectroscopic
  Properties of Rare Earth}}}\ (\bibinfo  {publisher} {John Wiley and Sons},\
  \bibinfo {address} {New York},\ \bibinfo {year} {1965})\BibitemShut {NoStop}%
\bibitem [{\citenamefont {Birgeneau}(1967)}]{BirgeneauRJ67}%
  \BibitemOpen
  \bibfield  {author} {\bibinfo {author} {\bibfnamefont {R.~J.}\ \bibnamefont
  {Birgeneau}},\ }\href@noop {} {\bibfield  {journal} {\bibinfo  {journal} {J.
  Phys. Chem. Solids}\ }\textbf {\bibinfo {volume} {28}},\ \bibinfo {pages}
  {2429} (\bibinfo {year} {1967})}\BibitemShut {NoStop}%
\bibitem [{\citenamefont {Buckmaster}(1962)}]{BuckmasterHA62}%
  \BibitemOpen
  \bibfield  {author} {\bibinfo {author} {\bibfnamefont {H.~A.}\ \bibnamefont
  {Buckmaster}},\ }\href@noop {} {\bibfield  {journal} {\bibinfo  {journal}
  {Can. J. Phys.}\ }\textbf {\bibinfo {volume} {40}},\ \bibinfo {pages} {1670}
  (\bibinfo {year} {1962})}\BibitemShut {NoStop}%
\end{thebibliography}
\end{document}